\documentclass[a4paper,11pt]{article}
\pdfoutput=1 

\usepackage{jheppub} 

\usepackage[T1]{fontenc} 
\usepackage{braket}
\usepackage{amsmath}
\usepackage{slashbox}
\usepackage{threeparttable}
\usepackage{graphicx}
\usepackage{subfigure}

\title{Toward the nonequilibrium thermodynamic analog of complexity and the Jarzynski identity}

\author[\dagger,*]{Chen Bai,}
\author[\dagger,*]{Wen-hao Li,}
\author[\dagger,*,\ddagger,1]{Xian-hui Ge\note{Corresponding author.}}

\affiliation[\dagger]{Department of Physics, College of Sciences, Shanghai University, Shanghai 200444, People's Republic of China}
\affiliation[*]{Shanghai Key Laboratory of High Temperature Superconductors, Shanghai University, Shanghai 200444, People's Republic of China}
\affiliation[\ddagger]{Center for Gravitation and Cosmology, Colleage of Physical Science and Technology, Yangzhou University, Yangzhou 225009, People's Republic of China}

\emailAdd{gexh@shu.edu.cn}
\emailAdd{chenbai@shu.edu.cn}
\emailAdd{3311980500@qq.com}

\abstract{The Jarzynski identity can describe small-scale nonequilibrium systems through stochastic thermodynamics. The identity considers fluctuating trajectories in a phase space. The complexity geometry frames the discussions on quantum computational complexity using the method of Riemannian geometry, which builds a bridge between optimal quantum circuits and classical geodesics in the space of unitary operators. Complexity geometry enables the application of the methods of classical physics to deal with pure quantum problems. By combining the two frameworks, i.e., the Jarzynski identity and complexity geometry, we derived a complexity analog of the Jarzynski identity using the complexity geometry. We considered a set of geodesics in the space of unitary operators instead of the trajectories in a phase space. The obtained complexity version of the Jarzynski identity strengthened the evidence for the existence of a well-defined resource theory of uncomplexity and presented an extensive discussion on the second law of complexity. Furthermore, analogous to the thermodynamic fluctuation-dissipation theorem, we proposed a version of the fluctuation-dissipation theorem for the complexity. Although this study does not focus on holographic fluctuations, we found that the results are surprisingly suitable for capturing their information. The results obtained using nonequilibrium methods may contribute to understand the nature of the complexity and study the features of the holographic fluctuations.}

\begin{document}
\maketitle
\flushbottom

\section{Introduction}
\label{sec:intro}
After Wheeler proposed the ``It from bit'' \cite{1}, an increasing number of concepts in information theory have been introduced into every corner of physics and have played important roles. One fascinating and novel example of such concepts is the quantum computational complexity\footnote{``Quantum computational complexity'' is referred to as         ``complexity''}, defined as the minimal number of primitive quantum gates required to generate a given unitary operator $U$:
\begin{equation}\label{1.1}
U=\underbrace{g_N...g_2g_1}_{\mathrm{Complexity}=N},
\end{equation}
where the fixed gate set $\{g_1,g_2,\dots,g_N\}$ comprises the primitive gates required to generate $U$. Complexity has been introduced as a theoretical tool for quantifying the difficulty faced in implementing a desired quantum computational task. It measures the hardness in constructing a given unitary operator $U$ (unitary complexity) or approximating a target quantum state from a reference state (state complexity).

In the context of AdS/CFT correspondence \cite{12,13,14}, several information quantities have natural duality in terms of geometric objects, which is considered to encode the features of holographic spacetime (e.g., entanglement entropy is dual to the area of extremal surfaces) \cite{15}. Similar to the entanglement entropy, complexity was recently conjectured to have a holographic dual. The two main holographic correspondences for complexity are ``\textit{Complexity=Volume}'' \cite{16,17,18} and ``\textit{Complexity=Action}'' \cite{19,20}. Chemissany and Osborne \cite{21} developed a procedure to directly associate a pseudo-Riemannian manifold with the dual AdS space (bulk spacetime) arising from a natural causal set induced by local quantum circuits. Additionally, they studied the fluctuations of the AdS space, which is caused by the dynamics of the dual boundary quantum system via \textit{the principle of minimal complexity}\footnote{This will be explained in the next section.}, and argued that the Brownian motion in the space of unitary operators might simulate such a fluctuation. Furthermore, they introduced a partition function by introducing a path integral in the space of unitary operators to capture information on holographic fluctuations.

To obtain a more quantitative comprehension of the complexity, a geometric treatment was proposed for the complexity in \cite{2} and solidified in \cite{3, 4, 5}. Based on previous works, a framework, called ``complexity geometry'' was gradually established \cite{6, 7}. We summarized the main idea of complexity geometry as follows: introduction of a Riemannian (or a Finsler) metric in the space of unitary operators\footnote{The group manifold, the space of unitary operators, and the configuration space in this article all refer to special unitary group with an introduced metric structure.} (note that the elements of the space of unitary operators act on a given number of qubits). Accordingly, the distance or action functional obtained from the metric is defined as the two measures of the complexity. Therefore, pure quantum (quantum scenario) complexity-related problems are changed into geometric problems that can be solved with classical mechanics (classical scenario) \cite{8}. In particular, geodesics in the space of unitary operators can be obtained by geodesic equations. Recently, complexity and its geometry have been used as an efficient tool to study extensive topics, such as the second law of complexity \cite{8}, black hole thermodynamics \cite{9}, the accelerated expansion of the universe \cite{10}, and quantum gravity \cite{11}.

In the past few decades, the study of nonequilibrium systems in high energy physics has become increasingly popular. Accordingly, several remarkable theoretical frameworks have been implemented to capture the features of nonequilibrium systems. One of the most eye-catching frameworks is the Jarzynski identity \cite{22, 23}, which connects equilibrium quantities with nonequilibrium processes. In particular, the Jarzynski identity builds a bridge between the equilibrium free energy difference, $\Delta F$, and the work done on the system during a non-equilibrium process, $W$. The Jarzynski identity is expressed in the following form:
\begin{equation}\label{1.2}
\left\langle \mathrm{exp}(- \beta W)\right\rangle=\mathrm{exp}(-\beta\Delta F),
\end{equation}
where $\beta$ denotes the inverse temperature. The bracket, $\left\langle\cdots\right\rangle$, represents the ensemble average of all possible values of $W$. There are several proofs of the Jarzynski identity \cite{24, 25, 26, 27, 28}. Hummer and Szabo proposed an elegant path integral proof \cite{27} of the Jarzynski identity based on the Feynman-Kac formula \cite{29, 30, 31, 32}. This has played a pivotal role in stochastic thermodynamics. Furthermore, as a novel tool, the Jarzynski identity has been used as diverse as renormalization group \cite{33}, Out-of-Time-Order correlators (OTOCs) \cite{34}, and the R$\acute{\mathrm{e}}$nyi entropy \cite{35} in holography and quantum information. Thus, it is natural to propose that the Jarzynski identity can be generalized to connect with another important information quantity, the complexity, which may provide us with deeper insights into high-energy physics.

The Jarzynski identity not only interrelates with the second law of thermodynamics but also characterizes a few fluctuation relations, including the fluctuation-dissipation theorem that significantly connects with entropy. Based on this, one of the core ideas we explored in this study was the derivation of a version of the Jarzynski identity for complexity using the path integral approach \cite{27}. In addition, we generalized the discussions in \cite{8} for systems with time-dependent Hamiltonians and derived a version of the fluctuation-dissipation relation for complexity. Because \cite{21} found that the fluctuations of a boundary quantum system are diametrically associated with those of a bulk spacetime, we suggested that the proposed fluctuation-dissipation theorem is a feasible tool for quantitatively exploring the holographic fluctuations\footnote{We will not delve into this issue or invoke any gravitational model in our paper. In this paper, we mainly focus on presenting the relationship between the complexity itself and the Jarzynski identity.}.

\begin{figure}[htbp]
\centering
\includegraphics[scale=0.36]{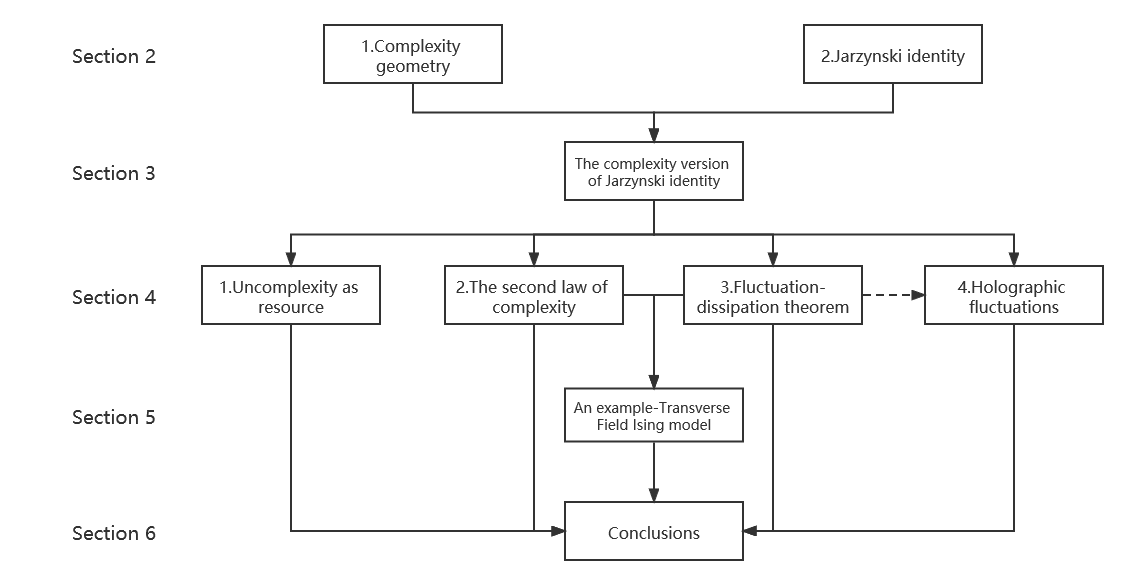} \label{P0}
\caption{Flowchart of the paper. The reason for connecting ``Fluctuation-dissipation theorem'' and ``Holographic fluctuations'' with a dashed line is that the former might be a potential tool to help us quantitatively study the latter.}
\end{figure}

In this study, we derived a complexity version of the Jarzynski identity, which is our main result. In addition, we argued that the obtained identity might bring us insights into several topics about complexity, in particular \textit{uncomplexity as a computational resource}, generalization of \textit{the second law of complexity}, complexity fluctuation-dissipation theorem, and holographic fluctuations. The remainder of this paper is organized as follows. In Section \ref{C2}, we briefly review the central concepts of the complexity geometry and the complexity version of the least action principle, namely \textit{the principle of minimal complexity}. We review a special derivation of the Jarzynski identity based on the path integral method. In Section \ref{C3}, we introduce the path integral in the space of unitary operators, and apply it to obtain the complexity version of the Jarzynski identity. The application of the Hamilton-Jacobi equation helps us rewrite the identity to a more intuitive form. In Section \ref{C4}, we discuss four issues about the complexity based on the obtained Jarzynski identity, which are the resource theory of uncomplexity, generalization of \textit{the second law of complexity} for stochastic auxiliary system $\mathcal{A}$, fluctuation-dissipation theorem in the context of quantum complexity, and holographic fluctuations. In Section \ref{C5}, we perform a numerical simulation of the transverse field Ising model to support our discussions on \textit{the second law of complexity}, where the complexity version of the Jarzynski identity plays a vital role. Finally, in Section \ref{C6}, we summarize our results and provide the conclusions and outlooks. This paper is structured in accordance with the flowchart presented in Fig. \ref{P0}.

\section{Preliminaries}\label{C2}
In this section, we first briefly review the notion of the complexity geometry. Two significant concepts are retrospected: the \textit{$\mathcal{Q}$-$\mathcal{A}$ correspondence} and \textit{the principle of minimal complexity}. Second, we present a review of the derivation of the Jarzynski identity presented in \cite{27}, which relies on the Feynman-Kac formula and path integral method. Finally, we generalize the derivation to adapt to our later discussions.

\subsection{Complexity geometry}\label{C21}
The complexity geometry is a powerful tool for quantifying the hardness to generate a specific unitary operator $U\in \mathrm{SU}(\mathrm{dim}[\mathcal{H}])$ from the identity $I\in \mathrm{SU}(\mathrm{dim}[\mathcal{H}])$, where ``$\mathrm{dim}[\mathcal{H}]$'' denotes the dimensions of the Hilbert space of the quantum systems (i.e., $\mathcal{H}$) comprising a fixed number of qubits.
\begin{figure}[htbp]
\centering
\includegraphics[scale=0.6]{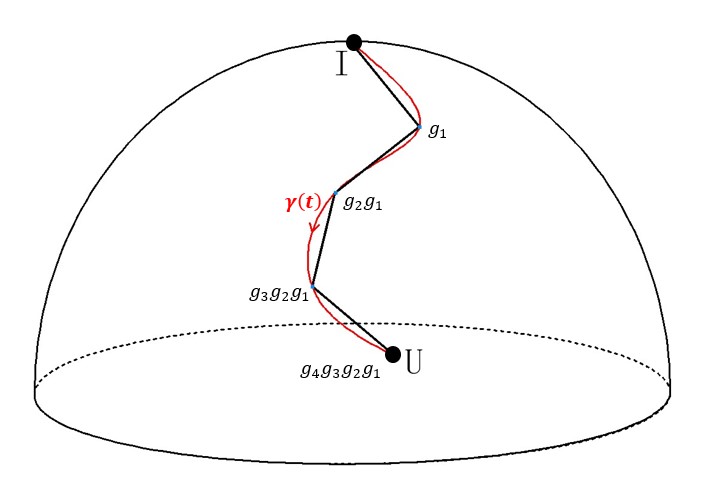} \label{P1}
\caption{The minimal number of polyline segments (black line) can be understood as the gate complexity, and the action of trajectory $\gamma(t)$ (red line) can be understood as a smooth representation of the complexity.}
\end{figure}

Instead of using gate complexity, we minimize a smooth function (the \textit{cost}) in a smooth manifold (the space of unitary operators) \cite{12}. Then, our purpose changes from how to determine the optimal quantum circuit comprising quantum gates to how to generate the target unitary operator
\begin{equation}\label{2.1}
U=\gamma(T)=\mathcal{P}e^{-i\int_0^T H(t) \mathrm{d}t}
\end{equation}
from a given Hamiltonian $H(t)$ with minimal \textit{cost} (the central idea is presented in Fig. 2) in a certain time interval $T$, where $\mathcal{P}$ represents the time-ordered operator. $\gamma(t)$ denotes the trajectory in the space of unitary operators. A \textit{cost} with fixed boundary conditions $\gamma(0)=I$, $\gamma(T)=U$ is defined as follows:
\begin{equation}\label{2.2}
A_a(U)\equiv \int_0^T L_a[\gamma(t),\dot{\gamma}(t)]\mathrm{d}t,
\end{equation}
where $L_a$ is a local functional of the $\gamma(t)\in SU(\mathrm{dim}[\mathcal{H}])$ in the space of unitary operators\footnote{The subscript ``$a$'' indicates an auxiliary system, which will be described in Section \ref{S212}.} that has the four following characteristics:
\begin{itemize}
\item[(1)]
\textit{Continuity}: $L_a[\gamma,\dot{\gamma}]\in C^{\infty}$.
\end{itemize}
\begin{itemize}
\item[(2)]
\textit{Non-negativity}: $L_a[\gamma,\dot{\gamma}]\ge0$ takes an equal sign if and only if $\dot{\gamma}=0$.
\end{itemize}
\begin{itemize}
\item[(3)]
\textit{Positive homogeneity}: $\forall\lambda\in\mathbb{R}$, $L_a[\gamma,\lambda\dot{\gamma}]=\lambda L_a[\gamma,\dot{\gamma}]$.
\end{itemize}
\begin{itemize}
\item[(4)]
\textit{Triangle inequality}: $\forall\dot{\gamma},\dot{\gamma}'$, $L_a$ satisfies the triangle inequality such that $L_a[\gamma,\dot{\gamma}]+L_a[\gamma,\dot{\gamma}']\ge L_a[\gamma,\dot{\gamma}+\dot{\gamma}']$.
\end{itemize}
If we regard $L_a$ in Eq. (\ref{2.2}) as a Lagrangian, $A_a(U)$ becomes the action functional of the trajectory connecting endpoints $I$ and $U$ in the space of unitary operators. Thus, complexity is defined as the minimal value of $A_a$,
\begin{equation}\label{2.3}
C(U)\equiv\underset{\gamma}{\mathrm{inf}}A_a(U)=\underset{\gamma}{\mathrm{inf}}\int_0^T L_a[\gamma(t),\dot{\gamma}(t)]\mathrm{d}t,
\end{equation}
and the infimum is over all possible trajectories. The four properties of $L_a$ define a smooth manifold $\mathcal{M}$\footnote{To be precise, the Finsler manifold is a type of generalization of the Riemannian manifold. See Chapter 8 of \cite{36} for the Finsler geometry.} equipped with a local metric, called the complexity metric,
\begin{equation}\label{2.4}
G_\gamma(\cdot , \cdot) : T_{\gamma} \mathcal{M} \times T_{\gamma} \mathcal{M}\to \mathbb{R},
\end{equation}
where $T_{\gamma} \mathcal{M}$ is the tangent space at $\gamma\in \mathcal{M}$. Note that in this study, $\mathcal{M}$ is nothing but the $\mathrm{SU}(\mathrm{dim}[\mathcal{H}])$ group manifold. We will use $\mathcal{M}$ to represent the group manifold $\mathrm{SU}(\mathrm{dim}[\mathcal{H}])$ (the space of unitary operators) throughout this study. More details can be found in \cite{2,3,4,5,6,7}.

\subsubsection{Real quantum system ``\textit{$\mathcal{Q}$}''}\label{S211}
The quantum system we considered comprises $K$ qubits. The interactions between these qubits are taken as $k$-local. Here, $k$-local means that the Hamiltonian of the system contains the interaction terms of not more than $k$ qubits. For example, a $2$-local Hamiltonian
\begin{equation}\label{2.5}
H(t)=\sum_{i,j}h_{ij}(t),
\end{equation}
where $h_{ij}(t)$ is a Hermitian operator acting on two arbitrary qubits $i$ and $j$. The general expression of a $k$-local Hamiltonian is
\begin{equation}\label{2.6}
H(t)=\sum_{i_1<i_2<...<i_k}\sum_{a_1=\{x,y,z\}}\cdots\sum_{a_k=\{x,y,z\}}J^{a_1,\cdots,a_k}_{i_1,\cdots,i_k}(t)\sigma_{i_1}^{a_1}\sigma_{i_2}^{a_2}\cdots\sigma_{i_k}^{a_k}.
\end{equation}
Schematically, it can be written as
\begin{equation}\label{2.7}
H(t)=J^M(t)\sigma_M(t),
\end{equation}
where $\sigma_I$ and $J^I(t)$ are the set of generalized Pauli matrices and coupling functions. $I$ runs over all $(4^K-1)$ Pauli matrices with corresponding nonzero couplings\footnote{$(4^K-1)$ is also the dimensions of $\mathrm{SU}(2^K)$ group.}. We follow the Einstein's summation convention here and thereafter. This time-dependent Hamiltonian generates Eq. (\ref{2.1}) and determines the dynamics of a quantum system.

What are the dynamics of the quantum system $\mathcal{Q}$? The system $\mathcal{Q}$ is a standard quantum system. Thus, when referring to its dynamics, we usually consider the evolution process of the states in its Hilbert space (the space of states) with $2^K$ dimensions. We choose a reference state $\ket{\Omega}$ and a target state $\ket{\Psi(T)}$ as the initial and final system states, respectively. We then define a moving point $\gamma(t)\in \mathcal{M}$. According to the Schr\"{o}dinger's picture, the evolution starting from the initial state at time $t=0$ to the final state at $t=T$ is achieved by applying a particular unitary operator $U=\gamma(T)$,
\begin{equation}\label{2.8}
\ket{\Psi(T)}=U\ket{\Omega},
\end{equation}
and the time-evolution (with constraints $\gamma(0)=I$ and $\gamma(T)=U$) of $\gamma(t)$ itself satisfies the Schr\"{o}dinger equation
\begin{equation}\label{2.9}
\frac{\mathrm{d}\gamma(t)}{\mathrm{d}t}=-iH(t)\gamma(t),
\end{equation}
where $H(t)$ is a traceless Hermitian operator, i.e., Eq. (\ref{2.7}).

In this study, we mainly focus on the time-dependent Hamiltonian form of Eq. (\ref{2.7}), which is a generalization of the case presented in \cite{8}. The only difference between these two is that the coupling $J$ in the present study varies with time but is constant in \cite{8}. Two typical examples with the time-independent Hamiltonian are the SYK model \cite{37,38,39,40} and thermofield-double (TFD) state \cite{41,42}.

\subsubsection{Auxiliary classical system ``\textit{$\mathcal{A}$}''}\label{S212}
Because of the application of the complexity geometry, we define a classical auxiliary system $\mathcal{A}$, along the lines of \cite{8}, as a system that describes the evolution of the unitary operators of the quantum system $\mathcal{Q}$. We must consider the following questions to define such a classical auxiliary system:
\begin{itemize}
\item[(1)]
What does system $\mathcal{A}$ look like?
\end{itemize}
\begin{itemize}
\item[(2)]
How do we define the distance (metric) in the configuration space of $\mathcal{A}$?
\end{itemize}
\begin{itemize}
\item[(3)]
What is the equation of motion?
\end{itemize}
Let us answer these questions one by one. The classical auxiliary system $\mathcal{A}$ describes the evolution of the unitary operators in $\mathcal{M}$; thus, the configuration space is the space of unitary operators $\mathcal{M}$ and each point in $\mathcal{M}$ corresponds to an element of the special unitary group. The number of degrees of freedom of system $\mathcal{A}$ is equal to the dimension of $\mathcal{M}$. Moreover, the evolution of a unitary operator in the space of unitary operators can be regarded as the motion of a fictitious nonrelativistic free particle with unit mass in the configuration space. The particle velocity is described by a tangent vector $\dot{\gamma}$ along the trajectory in $\mathcal{M}$. The tangent vector's dimension is consistent with the dimension of $\mathcal{M}$ (the number of degrees of freedom of system $\mathcal{A}$). For instance, consider a system $\mathcal{Q}$ comprising $K$ qubits and $\mathcal{M}=\mathrm{SU}(2^K)$. Then, the dual auxiliary system $\mathcal{A}$ has $(4^K-1)$ degrees of freedom and the tangent vectors of $\mathcal{M}$ are described by $(4^K-1)$-dimensional variables.

Because any Hermitian operator can be expanded in generalized Pauli matrices, such as Eq. (\ref{2.7}), we can consider the set of Pauli matrices as a set of basis in $\mathcal{M}$. The generalized Pauli matrices satisfy
\begin{equation}\label{2.10}
\mathrm{Tr}\sigma_M\sigma_N=\delta_{MN},
\end{equation}
where we assume that the trace ``$\mathrm{Tr}$'' is always normalized and $\delta_{MN}$ denotes the Kronecker-delta. Therefore, coupling $J^M(t)$ can be solved as
\begin{equation}\label{2.11}
J^M(t)=\delta^{MN}\mathrm{Tr}[i\dot{\gamma}(t)\gamma^{\dagger}(t)\sigma_N].
\end{equation}
If we set $\gamma(0)=I$, then at $t=0$ Eq. (\ref{2.11}) is in form of
\begin{equation}\label{2.12}
J^M(0)=\delta^{MN}\mathrm{Tr}[i\dot{\gamma}(t)\sigma_N] \vert_{t=0},
\end{equation}
where the right-hand side is the projection of the initial velocity onto the tangent space axes oriented along the Pauli basis. Brown and Susskind regarded $J^M(0)$ as the initial velocity of a fictitious particle of system $\mathcal{A}$ (i.e., $V^M(0)\equiv J^M(0)$). This is called the \textit{velocity-coupling correspondence} \cite{8}. Hence, $J^M(t)$ plays the role of a time-dependent velocity. Then, couplings can be written in terms of general (local) coordinates, that is, \{$J^M(t)\}\to \{\dot{X}^M(t)$\}, where \{$X^M(t)$\} denotes the coefficients (components) of some vectors in $\mathcal{M}$ expanded in the Pauli basis. Notably, the selection of local coordinates is not unique. An example of different choices is presented in \cite{43}.

Next, let us introduce the standard inner-product metric (bi-invariant) in $\mathcal{M}$, that is,
\begin{equation}\label{2.13}
\begin{aligned}
\mathrm{d}s^2|_{\mathrm{inner-product}}&=\mathrm{Tr}[\mathrm{d}U^{\dagger}\mathrm{d}U]\\
&=\delta_{MN}\mathrm{Tr}[iU^{\dagger}\mathrm{d}U\sigma^M]\mathrm{Tr}[iU^{\dagger}\mathrm{d}U\sigma^N],
\end{aligned}
\end{equation}
which equally treats all tangent directions $\sigma_I$. Mathematically, it means that if $\mathcal{M}$ is equipped with a bi-invariant metric, then $\mathcal{M}$ is homogeneous and isotropic. Such a bi-invariant metric induces the system $\mathcal{A}$ with characteristics similar to those in a classical system in the Euclidean space.

Recall that the complexity is a tool for measuring how difficult it is to generate a target unitary $U$. We generalize $\delta_{MN}$ to a symmetric positive-definite penalty factor $G_{MN}$ by extending Eq. (\ref{2.13}) to the complexity geometry condition \cite{6}. The metric then becomes
\begin{equation}\label{2.14}
\mathrm{d}s^2=G_{MN}\mathrm{Tr}[iU^{\dagger}\mathrm{d}U\sigma^M]\mathrm{Tr}[iU^{\dagger}\mathrm{d}U\sigma^N],
\end{equation}
which is a right-invariant local metric on $\mathcal{M}$. The metric is rewritten in terms of general coordinates as follows:
\begin{equation}\label{2.15}
\mathrm{d}s^2=G_{MN}\mathrm{d}X^M\mathrm{d}X^N.
\end{equation}
Eq. (\ref{2.15}) provides a homogeneous but anisotropic curved space (with negative curvature for a large number of qubits \cite{6,44}) as the configuration space. ``Anisotropic'' means that it is hard for a particle of the system $\mathcal{A}$ to move in some directions. In quantum computation, this means that it is tough to impose quantum gates in some directions to generate the unitary operator $U$, because these directions are severely penalized. A more detailed discussion can be found in \cite{6}. We mainly consider the (irreducible\footnote{In this study we consider that all Markov chains are irreducible.}) Markov processes in $\mathcal{M}$ in the following sections. The state-space $\mathcal{S}$ of these processes consists of the possible trajectories starting at the origin $I\in\mathcal{M}$ and has a fixed endpoint $U\in\mathcal{M}$, $\mathcal{S}\equiv\{\gamma_i\}$, where $i\in\{0,1,\cdots\cdots,N\}$ indicates each moment and satisfies $\gamma_0=\gamma(t_0)=I$ and $\gamma_N=\gamma(t_N)=U$. We can obtain the action functional of the trajectories in $\mathcal{M}$ using the metric presented in Eq. (\ref{2.15})
\begin{equation}\label{2.16}
A_a=\int  \frac{1}{2}G_{MN}\dot{X}^M\dot{X}^N\mathrm{d}t,
\end{equation}
where subscript ``$a$'' denotes a quantity of the system $\mathcal{A}$. Eq. (\ref{2.16}) is a rewritten form of Eq. (\ref{2.2}) after considering the Lagrangian
\begin{equation}\label{2.17}
L_a=\frac{1}{2}G_{MN}\dot{X}^M\dot{X}^N.
\end{equation}
Next, the complexity is calculated by minimizing $A_a$ in Eq. (\ref{2.16}) as
\begin{equation}\label{2.18}
C=\underset{\gamma}{\mathrm{inf}}\int\frac{1}{2}G_{MN}\dot{X}^M\dot{X}^N\mathrm{d}t,
\end{equation}
which is an explicit form of Eq. (\ref{2.3}).

For any classical system, the equation of motion can be derived from an action functional by applying the Euler-Lagrange equation. Thus, for any auxiliary system $\mathcal{A}$, the equation of motion reads
\begin{equation}\label{2.19}
 \frac{\partial L_a}{\partial X^M}-\frac{\mathrm{d}}{\mathrm{d} t}(\frac{\partial L_a}{\partial \dot{X}^M})=0.
\end{equation}
Substituting the right-hand side of Eq. (\ref{2.17}) into Eq. (\ref{2.19}), we obtain
\begin{equation}\label{2.20}
\ddot{X}^M+\Gamma^M_{YN}\dot{X}^Y\dot{X}^N=0,
\end{equation}
where the Christoffel symbol $\Gamma^M_{YN}$ is defined as
\begin{equation}\label{2.21}
\Gamma^M_{YN}\equiv \frac{1}{2}G^{MS}(\partial_NG_{SY}+\partial_YG_{SN}-\partial_SG_{NY}),
\end{equation}
and $\partial_M\equiv\frac{\partial}{\partial X^M}$. Eq. (\ref{2.20}) is the geodesic equation in $\mathcal{M}$, which is the equivalent expression to the equation of motion\footnote{There is one more equivalent form commonly used in Lie algebra, namely, the Euler-Arnold equation \cite{45}.}.

We further extend the discussion made by Brown and Susskind \cite{8}. Consider a system $\mathcal{Q}$ governed by a time-dependent Hamiltonian $H(t)$. Consequently, the dual auxiliary system $\mathcal{A}$ is a stochastic classical system, and the evolution of the unitary operator corresponds to a Markov process in $\mathcal{M}$. The equation of motion becomes a stochastic differential equation, e.g., \textit{Quantum Brownian Circuit} \cite{46},
\begin{equation}\label{2.22}
\mathrm{d}\gamma(t)=-\frac{1}{2}\gamma(t)\mathrm{d}t+\frac{i}{\sqrt{8K(K-1)}}\sum_{j<k}\sum_{\alpha_j,\alpha_k=0}^{3}\sigma_j^{\alpha_j}\otimes\sigma_k^{\alpha_k}\gamma(t)\mathrm{d}B_{j,k,\alpha_j,\alpha_k}(t),
\end{equation}
where $\mathrm{d}B_{j,k,\alpha_j,\alpha_k}(t)$ are independent Wiener processes with a unit variance per unit time. $K$ denotes the number of qubits in the quantum system $\mathcal{Q}$. The configuration space $\mathcal{M}$ is $\mathrm{SU}(2^K)$ (equipped with a complexity metric). The system $\mathcal{A}$ has $(4^K-1)$ degrees of freedom.

In summary, although the understanding of the complexity geometry is still incomplete, it helps us change pure quantum problems (i.e., finding the optimal circuits) to classical geometric problems (i.e., finding geodesics in $\mathcal{M}$). This quantum-classical duality is called the \textit{$\mathcal{Q}$-$\mathcal{A}$ correspondence}.

\subsubsection{Principle of minimal complexity}
\textit{The principle of minimal complexity} is the complexity version of the principle of least action, namely the application of the principle of least action to the auxiliary system $\mathcal{A}$ \cite{21}. The statement of \textit{the principle of minimal complexity} is the rewritten form of the principle of least action \cite{47}:
``A true dynamical trajectory of the system $\mathcal{A}$ between an initial and a final configuration in a specified time interval is found by imagining all possible trajectories that the system could conceivably take. Then, we compute the complexity for each of these trajectories and select one that makes the complexity stationary (or `minimal'). Thus, true trajectories are those that have the minimal complexity.''

By applying the variational method to the first order of Eq. (\ref{2.18}), the Euler-Lagrange equation for \textit{the principle of minimal complexity} can be obtained. However, Eq. (\ref{2.19}) is a necessary condition for the minimal value of complexity. To determine whether a trajectory is minimal, we must consider its second-order variation. A trajectory with minimal complexity has a positive second-order derivative.

A similar statement arises from the \textit{complexity=action} conjecture, that is, \textit{the principle of least computation} \cite{20}. The conjecture states that the complexity of the boundary state is proportional to the on-shell action of the bulk spacetime. Therefore, we can apply the principle of least action to obtain the equations of motion in the bulk spacetime and minimize the complexity.

\subsection{Jarzynski identity}\label{C22}
As one of the most remarkable achievements in recent decades, the Jarzynski identity can be derived or proved by various means such as microscopic \cite{22} or stochastic \cite{27} dynamics. In this study, our discussion is mainly based on the path integral derivation of the Jarzynski identity presented by Hummer and Szabo \cite{27}.

\subsubsection{Path integral derivation}
First, suppose there is a system whose phase space is denoted by $\vec{x}$. The evolution of the system follows the canonical Liouville equation, i.e.,
\begin{equation}\label{2.23}
\frac{\partial P(\vec{x},t)}{\partial t}=L_tP(\vec{x},t),
\end{equation}
where $P(\vec{x},t)$ is the phase space density function and $L_t$ is a time-dependent operator. Its stationary solution is a Boltzmann distribution $L_t\mathrm{e}^{-\beta H(\vec{x},t)}=0$ \cite{48}. Therefore, we consider a distribution $P(\vec{x},t)$ at time $t$ that satisfies the condition of the stationary solution $L_tP(\vec{x},t)=0$. At the same time, it obeys
\begin{equation}\label{2.24}
\frac{\partial P(\vec{x},t)}{\partial t}=-\beta\left(\frac{\partial H(\vec{x},t)}{\partial t}\right)P(\vec{x},t).
\end{equation}
If we combine the stationary solution condition of Eq. (\ref{2.23}) and Eq. (\ref{2.24}), a Fokker-Planck type equation with a sink term can be obtained, i.e.,
\begin{equation}\label{2.25}
\frac{\partial P(\vec{x},t)}{\partial t}=L_tP(\vec{x},t)-\beta\left(\frac{\partial  H(\vec{x},t)}{\partial t}\right)P(\vec{x},t).
\end{equation}
Now, we consider the system evolves from an equilibrium state at $t=0$ to a nonequilibrium state at $t=T$ under an arbitrary force. Under this condition, Hummer and Szabo determined that the solution of Eq. (\ref{2.25}) could be expressed using the Feynman-Kac formula \cite{29, 30, 31} as follows:
\begin{equation}\label{2.26}
P(\vec{x},T)=\left\langle{\delta(\vec{x}-\vec{x}(T))\mathrm{exp}{[-\beta\int_0^T\frac{\partial H}{\partial t}(\vec{x},t)\mathrm{d}t]}}\right\rangle,
\end{equation}
where the bracket $\langle{\cdots}\rangle$ represents the ensemble average. Each trajectory in the phase space is weighted by a factor that can be defined as the external work done on the system,
\begin{equation}\label{2.27}
W(T)\equiv \int_0^T\frac{\partial H(\vec{x},t)}{\partial t}\mathrm{d}t.
\end{equation}
Based on the famous relation between the free energy and partition function in statistical mechanics (i.e., $F(t)=-\beta^{-1}\mathrm{log}Z(t)$), the exponent of the free energy difference $\Delta F(T)=F(T)-F(0)$ is given as
\begin{equation}\label{2.28}
\mathrm{e}^{-\beta \Delta F(T)}=\frac{Z(T)}{Z(0)}=\frac{\int \mathrm{d}\vec{x} \mathrm{e}^{-\beta H(\vec{x},T)}}{\int \mathrm{d}\vec{y} \mathrm{e}^{-\beta H(\vec{y},0)}}.
\end{equation}
The Boltzmann distribution reads
\begin{equation}\label{2.29}
P(\vec{x},T)=\frac{\mathrm{e}^{-\beta H(\vec{x},T)}}{\int \mathrm{d}\vec{y} \mathrm{e}^{-\beta H(\vec{y},0)}}.
\end{equation}
In Eq. (\ref{2.29}), the numerator is divided by such a denominator because the initial distribution is exact. Thus the Jarzynski identity
\begin{equation}\label{2.30}
\mathrm{exp}(-\beta\Delta F(T))=\langle{\mathrm{exp}(-\beta W(T))}\rangle
\end{equation}
is derived by integrating both sides of Eq. (\ref{2.26}) over $\vec{x}$.

\subsubsection{Generalized form}
To generalize their formalism to a configuration space, we must re-interpret the meaning of each symbol in Eq. (\ref{2.23}). We re-interpret $\vec{x}$ and $P(\vec{x},t)$ as a random variable and distribution function, respectively. Then, the time-dependent operator $L_t$ becomes the Fokker-Planck operator that satisfies a Fokker-Planck equation \cite{49}. Consequently, the stationary solution of $L_t$ becomes Gaussian that can be expressed as
\begin{equation}\label{2.31}
L_t\mathrm{e}^{-\eta A}=0,
\end{equation}
where $\eta$ is a positive constant. The quadratic action $A$ is given by
\begin{equation}\label{2.32}
A={\frac{1}{2}}\int_0^T \delta_{ij}\dot{x}^i\dot{x}^j\mathrm{d}t,
\end{equation}
where $x^i$ and $x^j$ represent components of $\vec{x}$. If we consider a curved space, we only need to transform the Kronecker-delta $\delta_{ij}$ into a general metric tensor $g_{ij}$. Then, Eq. (\ref{2.24}) becomes
\begin{equation}\label{2.33}
\frac{\partial P(\vec{x},t)}{\partial t}=-\eta\left(\frac{\partial A(\vec{x},t)}{\partial t}\right)P(\vec{x},t).
\end{equation}
Similarly, combining this equation with Eq. (\ref{2.31}) yields
\begin{equation}\label{2.34}
\frac{\partial P(\vec{x},t)}{\partial t}=L_tP(\vec{x},t)-\eta\left(\frac{\partial  A(\vec{x},t)}{\partial t}\right)P(\vec{x},t).
\end{equation}
We apply the Feynman-Kac formula to Eq. (\ref{2.34}) to obtain
\begin{equation}\label{2.35}
P(\vec{x},T)=\left\langle{\delta(\vec{x}-\vec{x}(T))\mathrm{exp}{[-\eta\int_0^T\frac{\partial A}{\partial t}(\vec{x},t)\mathrm{d}t]}}\right\rangle.
\end{equation}
We can further define the generalized work\footnote{We sometimes refer to the action functional as the energy functional, which is why the generalized work is defined in this way, e.g., \cite{33}.} as
\begin{equation}\label{2.36}
W(t)\equiv \int_0^t\frac{\partial A}{\partial s}(\vec{x},s)\mathrm{d}s.
\end{equation}
By integrating $\vec{x}$ on both sides of Eq. (\ref{2.35}), we obtain a Jarzynski-like identity, that is,
\begin{equation}\label{2.37}
\frac{\int D\vec{x}\mathrm{e}^{-\eta A(\vec{x},T)}}{\int D\vec{y}\mathrm{e}^{-\eta A(\vec{y},0)}}\equiv\frac{Z(T)}{Z(0)}=\langle{\mathrm{exp}(-\eta W(T))}\rangle,
\end{equation}
where $Z(t)$ is the partition function and $D\vec{x}$ and $D\vec{y}$ are suitable path integral measures in the configuration space. Note that if we regard the constant $\eta$ as the inverse temperature of the system and set $T=1$, using the $F(t)=-\eta^{-1}\mathrm{log}Z(t)$ relation, Eq. (\ref{2.37}) becomes the Jarzynski identity (i.e., Eq. (\ref{2.30})).

To apply this generalized method to the space of unitary operators, we will introduce the Haar measure and ergodicity in the subsequent section. Furthermore, we will use the Hamilton-Jacobi (HJ) equation to rewrite the complexity version of the Jarzynski identity.

\section{Jarzynski identity under the background of the complexity geometry}\label{C3}
In Section \ref{C31}, we introduce the Haar measure and path integral in $\mathcal{M}$. Then, we apply the same logic as in the path integral derivation in the last section to obtain the Jarzynski identity for complexity. We argue that one can use the Hamilton-Jacobi equation to rewrite the obtained identity into a more meaningful form and raise several nonequilibrium analogs of complexity dynamical issues. We will present these in the next section.

\subsection{Path integral in $\mathcal{M}$}\label{C31}
\subsubsection{Path integral}\label{C311}
Before discussing the path integral in $\mathcal{M}$ (mathematically, the path integral is somewhat consistent with the Wiener measure; see \cite{50} for the definition of the Wiener measure), we must first clarify a prerequisite that enables to integrate in the group manifold. We define a Haar measure as a unique measure that is invariant under translations by group elements. We express a Haar measure in terms of general coordinates:
\begin{equation}\label{3.1}
[\mathrm{d}\gamma]\equiv \frac{1}{N_c}\sqrt{G_{MN}(\gamma)}\mathrm{d}X^1\mathrm{d}X^2\cdots \mathrm{d}X^{\mathrm{dim}(\mathcal{M})}.
\end{equation}
We use $[\mathrm{d}\gamma]$ to represent the Haar measure (see \cite{43} for an example for choosing a specific parameterization of a normalized Haar measure). $N_c$ is the normalization coefficient. $\sqrt{G_{MN}(\gamma)}\equiv |\mathrm{det}(J_d)|$ denotes the determinant of the Jacobian matrix $J_d$. The Jacobian matrix identifies an invariant measure under coordinate transformations. We follow the notation in Eq. (\ref{2.15}), i.e., $\{X^M\}$ are the coefficients (or components) of $\gamma(t)\in\mathcal{M}$ expanded on a local basis in all the following contents. The Haar measure must meet two requirements:
\begin{itemize}
\item[(1)]
normalization condition: $\int_{\mathcal{M}} [\mathrm{d}\gamma]=I$; and
\end{itemize}
\begin{itemize}
\item[(2)]
orthogonal completeness condition: $\int_{\mathcal{M}}[\mathrm{d}\gamma]\ket{\gamma}\bra{\gamma}=I$.
\end{itemize}
Here, $\ket{\gamma}$ is considered as the group representation of some \textit{pseudo-quantum states} that satisfies $\langle\gamma|\gamma'\rangle=\delta(\gamma-\gamma')$. Note that the system described by these \textit{pseudo-quantum states} is not a real quantum system, but a hypothetical quantum system obtained by applying ``\textit{stochastic quantization}'' \cite{51} to system $\mathcal{A}$, a classical system.

Note that the \textit{pseudo-quantum states} only help us derive the path integral. To derive the path integral in quantum mechanics, the evolution kernel is obtained by constantly inserting the orthogonal completeness condition, which is familiar to most physicists. Thus, we can assume that there exists a hypothetical quantum system with such an orthogonal completeness condition. Then, we can derive the path integral by inserting the orthogonal completeness condition instead of introducing an unfamiliar concept, ``i.e., \textit{stochastic quantization}'' \cite{51}. In this system, a stochastic differential equation, e.g., Langevin equation, is analogous to the Heisenberg operator equation in quantum mechanics. Although the introduction of the \textit{pseudo-quantum state} is not mathematically rigorous, it is convenient for us to introduce a path integral in $\mathcal{M}$. A more rigorous discussion of ``\textit{stochastic quantization}'' can be found in \cite{51}. It must be emphasized that the hypothetical quantum system is essentially a classical stochastic system, and its uncertainty comes from stochastic motion not from the Heisenberg uncertainty principle. Therefore, the system described here is purely classical even if we use Dirac notations.

Once a Haar measure is defined, we can do integral in $\mathcal{M}$. Thus, we can derive the evolution kernel. Consider a Markov chain in $\mathcal{M}$ with a finite state space (i.e., $\mathcal{S}=\{\gamma_0=I,\gamma_1,\cdots\cdots,\gamma_N=U\}$ with $0=t_0<t_1<t_2<\cdots\cdots <t_{N-1}<t_N=T$). We set the unit time interval as $t_i-t_{i-1}=\Delta t=T/N$ for any $i\in \{1,2,\cdots\cdots,N\}$. Thus, the propagator $\mathcal{K}_a(\gamma_{i+1},t_{i+1};\gamma_i,t_i)$ is defined as
\begin{equation}\label{3.2}
\mathcal{K}_a{(\gamma_{i+1},t_{i+1};\gamma_i,t_i)} \equiv \braket{\gamma_{i+1},t_{i+1}|\gamma_i,t_i}=e^{iL_a[\gamma(t_{i+1});\gamma(t_i)]\Delta t+O(\Delta t^2)}.
\end{equation}
We can obtain the evolution kernel by repeatedly inserting the orthogonal completeness condition similar to what we usually do in Feynman path integrals. If we take N $\to \infty$ such that $\Delta t \to 0$ , the evolution kernel (or heat kernel) $K_a(U,T;I,0)$ from $I$ at $t=0$ to $U$ at $t=T$ is obtained as
\begin{equation}\label{3.3}
\begin{aligned}
K_a(U,T;I,0)&\equiv \int_{\mathcal{M}} \prod_{i=0}^{N-1}[\mathrm{d}\gamma_i]\mathcal{K}_a{(\gamma_{i+1},t_{i+1};\gamma_i,t_i)}\\ &= \prod_{i=0}^{N-1}K_a(\gamma_{i+1},t_{i+1};\gamma_i,t_i).
\end{aligned}
\end{equation}
The sum of Lagrangians, $L_a$, can be written in the form of the complexity
\begin{equation}\label{3.4}
K_a(U,T;I,0)=\int_{\mathcal{M}} \prod_{i=0}^{N-1}[\mathrm{d}\gamma_i]\mathrm{e}^{iC(T)}.
\end{equation}
However, the path integral in a curved configuration space inevitably introduces a correction term with a scalar curvature in the complexity $C$ to maintain the covariance of the path integral under any coordinate transformation. Consequently, we must deal with an additional curvature term when calculating the path integral.

Fortunately, an elegant proposal \cite{52} provides a novel form without any curvature modification; it introduces a factor called Van Vleck-Morette determinant \cite{53, 54} in Eq. (\ref{3.4})
\begin{equation}\label{3.5}
K_a(U,T;I,0)=\int_{\mathcal{M}} \prod_{i=1}^{N-1}[\mathrm{d}\gamma_i]\frac{N_c{}_i|\Delta(\gamma_*;\gamma_i)|}{\sqrt{G_{MN}(\gamma_*)}}\mathrm{e}^{iC(T)},
\end{equation}
where $N_c{}_i$ denotes the normalized factor of the $i$th Haar measure and $|\Delta(\gamma_*;\gamma)|$ is the Van Vleck-Morette determinant:
\begin{equation}\label{3.6}
|\Delta(\gamma_*;\gamma)|\equiv \frac{N_c^2}{\sqrt{G_{MN}(\gamma_*)}\sqrt{G_{MN}(\gamma)}}\mathrm{det}\left(-\frac{\delta^2\lambda(\gamma_*;\gamma)}{\delta X^M\delta X_*^N}\right),
\end{equation}
where $\lambda(\gamma_*;\gamma)$ is defined as the geodesic interval \cite{55, 56} between a fixed point $\gamma_* \in \mathcal{M}$ and $\gamma\in \mathcal{M}$. Note that
\begin{equation}\label{3.7}
\lambda(\gamma_*;\gamma)\equiv \frac{1}{2}D^2(\gamma_*;\gamma),
\end{equation}
where $D(\gamma_{*};\gamma)$ is the length of the geodesic connecting point $\gamma_{*}$ to point $\gamma$. With the replacement, i.e., $\int_{\mathcal{M}} D\gamma\equiv\int_{\mathcal{M}} \prod_{i=1}^{N-1}[\mathrm{d}\gamma_{i}]\frac{N_{c_i}|\Delta(\gamma_{*};\gamma_{i})|}{\sqrt{G_{MN}(\gamma_{*})}}$, the expression of evolution kernel becomes
\begin{equation}\label{3.8}
K_a(U,T;I,0)=\int_{\mathcal{M}} D\gamma\mathrm{e}^{iC(T)}.
\end{equation}
The detailed derivation is presented in Appendix \ref{F2} (see also \cite{52}). In time limit $t \to \infty$ we consider the continuation of time $t$ to the complex plane. A Wick rotation, $t \to i\eta t$, is then applied to Eq. (\ref{3.8}) such that one can rewrite the evolution kernel as
\begin{equation}\label{3.9}
Z_a(T)=\int_{\mathcal{M}} D\gamma\mathrm{e}^{-\eta C(T)},
\end{equation}
where $Z_a(T)$ refers to the partition function of the system $\mathcal{A}$ and $\eta$ is a positive constant\footnote{It has two meanings, a Lagrangian multiplier and the inverse temperature of the system $\mathcal{A}$ \cite{8}.}. Let us give some remarks on this equation. In principle, one can consider a trajectory with a large complexity in $\mathcal{M}$. Then, the contribution of this trajectory to Eq. (\ref{3.9}) is incredibly small because the complexity exists in the form of an exponential function, specifically, $e^{-\eta C}$, in Eq. (\ref{3.9}). We will further consolidate this fact in Appendix \ref{F3} by discussing the relationship between \textit{the principle of minimal complexity} and \textit{the second law of complexity}.

\subsubsection{Ergodicity}\label{C312}
The ergodic motion in the configuration space ensures that the contributions of all possible trajectories are included in Eq. (\ref{3.9}). The ergodicity for the system $\mathcal{A}$ can be fulfilled in two ways: partial ergodic (chaotic evolution with a time-independent Hamiltonian \cite{8}) and complete ergodic (stochastic process with a time-dependent Hamiltonian \cite{21}).

A time-independent $k$-local Hamiltonian generates the partial ergodic motion in $\mathcal{M}$. To understand this, consider a quantum system comprising $K$ qubits that evolves by applying the unitary operator
\begin{equation}\label{3.10}
U=\mathrm{e}^{-iHT}=\sum_{n=1}^{2^K}\mathrm{e}^{-iE_nT}\ket{E_n}\bra{E_n},
\end{equation}
where $\ket{E_n}$ are the eigenstates of Hamiltonian $H$ with eigenvalues $E_n$. Because ergodicity is equivalent to the incommensurability of the energy eigenvalues in this case and there are $2^K$ energy eigenvalues for the Hamiltonian $H$, the unitary operator $U$ moves on a $2^K$ dimensional torus (subspace of $\mathcal{M}$) \cite{8} in an ergodic motion.

Complete ergodicity can be achieved in the case with a time-dependent $k$-local Hamiltonian, i.e., Eq. (\ref{2.6}). In this case, the motions in $\mathcal{M}$ are considered as the ergodic Markov process filling up all $(4^K-1)$ dimensions of $\mathcal{M}$. An ergodic Markov process must rigorously satisfy irreducible and non-periodic conditions, and all states are persistent \cite{57}. The necessary and sufficient condition for the existence of a stationary distribution of an irreducible Markov chain is an ergodic Markov chain. We can write a stationary distribution according to our settings as follows:
\begin{equation}\label{3.11}
P(\gamma,t)=\mathcal{N}_a\mathrm{exp}(-\eta C(t)),
\end{equation}
where $\mathcal{N}_a$ represents the normalized constant. Thus, the ergodicity can be satisfied. In conclusion, the existence of a stationary distribution indicates that the motion of the unitary operator in $\mathcal{M}$ is ergodic when system $\mathcal{Q}$ is governed by a time-dependent Hamiltonian.

\subsection{Complexity version of the Jarzynski identity}\label{C32}
\subsubsection{Derivation of the Jarzynski identity}\label{C321}
In this section, we derive the complexity version of the Jarzynski identity from the Fokker-Planck equation with a sink term in $\mathcal{M}$. Consider a stochastic auxiliary system $\mathcal{A}$ that describes a particle moving from a fixed point $\gamma(0)\in \mathcal{M}$ to another fixed point $\gamma(T)\in \mathcal{M}$. The dual quantum system $\mathcal{Q}$ is governed by a time-dependent Hamiltonian $H(t)$ evolving in a time interval $T$. Here, the fixed endpoints of the trajectories in $\mathcal{M}$ play a similar role to the two fixed states in a common Jarzynski case. The evolution equation of the system $\mathcal{A}$ is a Fokker-Planck equation, i.e.,
\begin{equation}\label{3.12}
\frac{\partial P(\gamma,t)}{\partial t}=L_tP(\gamma,t),
\end{equation}
where $P(\gamma,t)$ represents the distribution function, and the time-dependent operator $L_t$ denotes the Fokker-Planck operator\footnote{The derivations of the Fokker-Planck equation are presented in Appendix \ref{F2}.}. Because of Eq. (\ref{3.11}) and Eq. (\ref{3.9}), we can construct a distribution as the stationary solution of Eq. (\ref{3.12}) at $t=T$ similar to what we did in Section \ref{C22}:
\begin{equation}\label{3.13}
P(\gamma,T)=\frac{1}{Z_a(0)}\mathrm{exp}(-\eta C(T)),
\end{equation}
where $Z_a(0)$ refers to the partition function at $t=0$ and complexity $C$ plays the role of the action $A$ in Section \ref{C22}, such that $P(\gamma,t)$ satisfies $L_tP(\gamma,t)=0$. Moreover, one can check that
\begin{equation}\label{3.14}
\frac{\partial P(\gamma,t)}{\partial t}=-\eta \left(\frac{\partial C(t)}{\partial t}\right) P(\gamma,t).
\end{equation}
Hence, using this equation and $L_tP(\gamma,t)=0$, we can obtain the Fokker-Planck equation with a sink term
\begin{equation}\label{3.15}
\frac{\partial P(\gamma,t)}{\partial t}=L_tP(\gamma,t)-\eta \left(\frac{\partial C(t)}{\partial t}\right) P(\gamma,t).
\end{equation}
Solving this equation with constraint $\gamma(T)=\gamma$, we obtain
\begin{equation}\label{3.16}
P(\gamma,T)=\left\langle{\delta(\gamma-\gamma(T))\mathrm{exp}{[-\eta\int_0^T\frac{\partial C}{\partial t}(\gamma,t)\mathrm{d}t]}}\right\rangle.
\end{equation}
The ensemble average is over all the possible trajectories departing from the identity $I$ to reach the fixed point $U$ at $t=T$. The Dirac function indicates the termination condition. Equating this equation with Eq. (\ref{3.13}) gives:
\begin{equation}\label{3.17}
\frac{1}{Z_a(0)}\mathrm{exp}(-\eta C(T))
=\left\langle{\delta(\gamma-\gamma(T))\mathrm{exp}{[-\eta\int_0^T\frac{\partial C}{\partial t}(\gamma,t)\mathrm{d}t]}}\right\rangle.
\end{equation}
By integrating $\gamma$ on both sides of this equality and defining a quantity, called computational work $W_a(t)$ as a type of general work defined in Eq. (\ref{2.36}) with the following form:
\begin{equation}\label{3.18}
W_a(t)\equiv \int_0^t\frac{\partial C}{\partial s}(\gamma,s)\mathrm{d}s,
\end{equation}
the complexity version of Jarzynski identity is given as
\begin{equation}\label{3.19}
\frac{Z_a(T)}{Z_a(0)}=\left\langle \mathrm{exp}(-\eta W_a(T))\right\rangle,
\end{equation}
which is one of our main proposals in this paper.

To simplify Eq. (\ref{3.19}), we introduce an analog of the thermodynamic free energy in complexity, that is, the ``computational free energy'':
\begin{equation}\label{3.20}
F_a(t)\equiv -\eta^{-1}\mathrm{log}Z_a(t).
\end{equation}
If we assume $\eta=1/T_a$ as the inverse temperature of the system $\mathcal{A}$ and set $t=1$, $F_a$ can be regarded as the thermodynamic free energy of system $\mathcal{A}$ and $Z_a(t)$ takes the same form as the partition function of a free particle\footnote{Essentially, this is the relationship between the partition function obtained from the path integral approach and the thermodynamic free energy \cite{58}.}.
A similar discussion between the complexity-related and thermodynamic quantities was made in \cite{8}. By substituting Eq. (\ref{3.20}) into Eq. (\ref{3.19}), the equality takes a more familiar form:
\begin{equation}\label{3.21}
\mathrm{exp}(-\eta \Delta F_a(T))=\left\langle \mathrm{exp}(-\eta W_a(T))\right\rangle,
\end{equation}
where $\Delta F_a(T)=F_a(T)-F_a(0)$ depends on the two endpoints of evolution in system $\mathcal{A}$.

Even though we have already defined the computational work $W_a$, intuitively understanding its physical meaning remains hard. Hence, we expect a more instructive interpretation of $W_a$ exists within the abovementioned discussion of the complexity version of the Jarzynski identity. Eq. (\ref{3.18}) suggests that the definition of $W_a$ contains the time derivative of complexity. We intend to rewrite the complexity version of the Jarzynski identity using the Hamilton-Jacobi (HJ) equation that describes the change of complexity.

\subsubsection{Hamilton-Jacobi equation}\label{C322}
To further explore the Jarzynski identity in the context of complexity, a proper rewriting of the expression is required. In this section, we start from the derivation of the HJ equation considering the complexity geometry and use the HJ equation to rewrite Eq. (\ref{3.21}). The rewriting of the Jarzynski identity will facilitate the development of the thermodynamic analog of complexity, particularly for \textit{the second law of complexity} and the related topic \textit{uncomplexity as a resource} \cite{8}.

Recall that the trajectories in $\mathcal{M}$ follow \textit{the principle of minimum complexity}, that is,
\begin{equation}\label{3.22}
\delta C(\gamma,\dot{\gamma})=\delta \int_0^TL_a(\gamma,\dot{\gamma},t)\mathrm{d}t=0.
\end{equation}
\begin{figure}[htbp]
\centering
\includegraphics[scale=0.65]{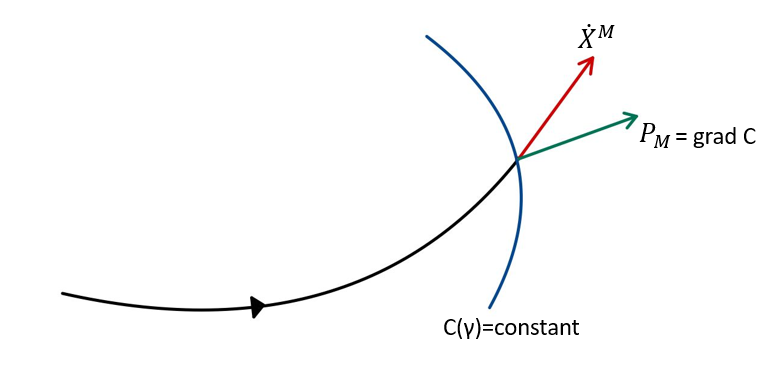} \label{P31}
\caption{The red arrow represents the direction of motion of a particle in the auxiliary system (tangent to the trajectory $\gamma(t)$), the green arrow represents the direction of the particle's momentum (perpendicular to the blue curve), and the blue curve represents a surface of equal complexity.}
\end{figure}
Now, let us consider \textit{the principle of minimal complexity} from a different perspective, that is, the Hamiltonian mechanics. We first determine the starting and ending points on a configuration space (i.e., $(\gamma(0)=I,t=0)$ and $(\gamma(T)=U,t=T)$), respectively. Next, we assume that the trajectories connecting those points are obtained using Eq. (\ref{3.22}), which satisfies Eq. (\ref{2.19}). The generalized momentum is defined as
\begin{equation}\label{3.23}
P_M\equiv \frac{\partial L_a}{\partial \dot{X}^M},
\end{equation}
whose direction is depicited in Fig. 3. Next, we rewrite Eq. (\ref{3.22}) as
\begin{equation}\label{3.24}
\delta C=\int_0^T\left\{\frac{\partial L_a}{\partial \dot{X}^M}\delta \dot{X}^M+\frac{\partial L_a}{\partial X^M}\delta X^M\right\}\mathrm{d}t.
\end{equation}
By substituting Eq. (\ref{2.19}) into this equation and assuming a small variation at the endpoint, namely $\delta \gamma(T)=\delta \gamma \not= 0$, Eq. (\ref{3.22}) is reformulated as
\begin{equation}\label{3.25}
\begin{aligned}
\delta C &=\int_0^T\left\{\frac{\partial L_a}{\partial \dot{X}^M}\delta \dot{X}^M+\frac{\mathrm{d}}{\mathrm{d}t}(\frac{\partial L_a}{\partial \dot{X}^M})\delta X^M\right\}\mathrm{d}t\\
&=\int_0^T\frac{d}{dt}\left\{\frac{\partial L_a}{\partial \dot{X}^M}\delta X^M\right\}\mathrm{d}t\\
&=P_M\delta X^M.
\end{aligned}
\end{equation}
We take the limit $\delta \gamma\to 0$\footnote{In Lagrangian mechanics, we always assume the variations of endpoints to be zero when deriving the Euler-Lagrange equation. Why do we need to consider an infinitesimal nonzero variation at the endpoint when deriving the HJ equation, though the Euler-Lagrange equation and the HJ equation can be used to describe the same classical system? The difference comes from different essential settings. In Lagrangian mechanics, we identify the equation of motions by varying trajectories between two fixed endpoints. However, when deriving the HJ equation, we consider that the trajectory satisfies the equation of motion. Therefore, instead of varying trajectories, we infinitesimally vary the endpoint of the trajectory to study the corresponding change of the action. So does for the same case in complexity story. In particular, \cite{59,60} have investigated a specific dynamic law, i.e., \textit{the first law of complexity} by studying the variation of a trajectory's endpoint in the complexity geometry.}, such that $P_M=\frac{\partial C}{\partial X^M}$. Thus, the complexity can be regarded as a functional of $\gamma$, and its infinitesimal variation is written as
\begin{equation}\label{3.26}
\delta C=\frac{\partial C}{\partial X^M} \delta X^M+\frac{\partial C}{\partial t}\mathrm{d}t.
\end{equation}
Dividing its both sides by $\mathrm{d}t$ and assuming that $\frac{\mathrm{d}C}{\mathrm{d}t}= L_a$ we obtain
\begin{equation}\label{3.27}
\frac{\partial C}{\partial t}=L_a-P_M\dot{X}^M=-H_a,
\end{equation}
where $H_a$ is the Hamiltonian of the system $\mathcal{A}$ and Eq. (\ref{3.27}) represents the HJ equation in the complexity story. However, a new question arises: what is the form of $H_a$? We look at the auxiliary Lagrangian, i.e., Eq. (\ref{2.17}), in which $P_M$ has the following form:
\begin{equation}\label{3.28}
\begin{aligned}
P_M&=\frac{\partial C}{\partial X^M}\\
&=\int_0^T\frac{\partial L_a}{\partial X^M}dt\\ &=\int_0^T\frac{\mathrm{d}}{\mathrm{d}t}\left(\frac{\partial L_a}{\partial \dot{X}^M}\right)\mathrm{d}t\\
&=G_{MN}\dot{X}^N.
\end{aligned}
\end{equation}
The last equal sign is established because the metric tensor $G_{MN}$ is Hessian \cite{2}, that is, $G_{MN}\equiv \frac{\partial L_a}{\partial \dot{X}^M \partial \dot{X}^N}$. Based on the above construction, we have
\begin{equation}\label{3.29}
H_a=L_a=\frac{1}{2}G_{MN}\dot{X}^M\dot{X}^N,
\end{equation}
for which we have
\begin{equation}\label{3.30}
\frac{\partial C}{\partial t}+L_a=0.
\end{equation}
Substituting this into Eq. (\ref{3.18}), the computational work is recast as
\begin{equation}\label{3.31}
W_a(T)=-\int_0^TL_a(t)\mathrm{d}t=-C(T).
\end{equation}
We then obtain the equivalent expression of the Jarzynski identity as
\begin{equation}\label{3.32}
\mathrm{exp}(-\eta\Delta F_a(T))=\left\langle \mathrm{exp}(\eta C(T))\right\rangle.
\end{equation}
This equality directly builds a bridge between the computational free energy difference and the complexity.

In the next section, we will argue that the thermodynamic analogs of complexity should be explored based on Eq. (\ref{3.32}), because complexity has similarities with the thermodynamic entropy \cite{8} and the Jarzynski identity strongly connects with the entropy.

\section{On the nonequilibrium thermodynamic analog of complexity}\label{C4}
The Jarzynski identity can provide a theoretical framework for exploring the thermodynamics of nonequilibrium systems, including stochastic systems. Because we have derived the Jarzynski identity, Eq. (\ref{3.32}), and the system $\mathcal{A}$ is stochastic, this section primarily aims to construct the thermodynamic analogs and deepen our understanding of complexity. In addition, several interesting issues are discussed in this section. First, we will review the content and development for each topic. Next, we will further explore these topics based on the previously obtained results.

\subsection{Uncomplexity as a computational resource}
To understand this statement, we must first know what a resource is. The resource theory has a wide range of applications in quantum physics \cite{61,62,63,64,65,66}, and we do not need to know all about them. All we need to learn from the resource theory in this paper can be summarized by the following sentence: a resource is something one needs to do X \cite{67}. For example, negentropy is a resource needed for doing work \cite{61,62,63}, which is defined as the difference between the maximal and actual entropies,
\begin{equation}\label{4.2}
\mathrm{Negentropy}(t)\equiv S_{\mathrm{max}}-S(t).
\end{equation}
The system that does work (to achieve some goals) must expend negentropy. Therefore, negentropy is a resource for doing work. Because complexity shows its analogs with classical entropies \cite{8}, by analogy, a complexity version of negentropy, namely uncomplexity, is defined as
\begin{equation}\label{4.2}
\mathrm{Uncomplexity}(t)\equiv C_{\mathrm{max}}-C(t),
\end{equation}
where $C_{\mathrm{max}}$ is the possible maximal complexity and $C(t)$ denotes the actual complexity of the system $\mathcal{Q}$ at a certain moment. This quantity is a resource that can be expended for doing direct computations \cite{8,67}. The central idea is expressed as
\begin{equation}\label{4.3}
F_a\propto -C,
\end{equation}
where $F_a$ refers to the thermodynamic free energy of system $\mathcal{A}$ \cite{8,67} obtained by assuming $\eta=1/T_a$ as the inverse temperature of the system $\mathcal{A}$ and setting $T=1$ in Eq. (\ref{3.20}). Equivalently,
\begin{equation}\label{4.4}
R(t)\equiv C_{\mathrm{max}} -C(t),
\end{equation}
where resource is denoted by $R(t)$.

Suppose that a particle is initialized at $I\in \mathcal{M}$ with zero complexity in the system $\mathcal{A}$. Equivalently, no gate is initially applied to any qubit in the system $\mathcal{Q}$. Recall that uncomplexity is the space for complexity to grow \cite{8}. We can write
\begin{equation}\label{4.5}
F_a(0)=R(0)=C_{\mathrm{max}},
\end{equation}
because $F_a$ represents the ability of the system $\mathcal{Q}$ to do computation\footnote{$-\Delta F_a$ always takes non-negative values because $F(0)=C_{\mathrm{max}}\ge F_a(T)$.} (analogous to the case in thermodynamics). Substituting Eq. (\ref{4.4}) and Eq. (\ref{4.5}) into Eq. (\ref{3.32}), we obtain
\begin{equation}\label{4.6}
\mathrm{exp}(-\eta F_a(T))=\left\langle \mathrm{exp}(-\eta R(T))\right\rangle=\left\langle \mathrm{exp}\left[-\eta (C_{\mathrm{max}}-C(T))\right]\right\rangle.
\end{equation}
This equation provides new evidence supporting the existence of a well-defined resource theory of uncomplexity \cite{87}.

\subsection{Second law of complexity and fluctuation theorem}\label{C42}
\textit{The second law of complexity} is obtained by applying the thermodynamic method to the auxiliary system $\mathcal{A}$ and has been studied when the system $\mathcal{A}$ is chaotic \cite{8}. Based on the same logic, we should use stochastic thermodynamics to study \textit{the second law of complexity} when the system $\mathcal{A}$ is stochastic rather than chaotic. We need two important pieces to complete this ``puzzle'': \textit{the second law of thermodynamics} and \textit{the trajectory thermodynamics} \cite{28}. In Section \ref{C421}, we will first review these two important pieces in the framework of nonequilibrium thermodynamics. Subsequently, in Section \ref{C422}, we will discuss \textit{the second law of complexity} for stochastic auxiliary systems using an analogy with discussions of Section \ref{C421}. To avoid confusion, we specify a few significant notations and make a clarification before the discussion.

\textit{Notations}: in the following content, we use
\begin{equation}\label{4.7}
P(\gamma)\equiv P(\gamma,T)=\frac{e^{-\eta C(\gamma)}}{Z_a(0)}
\end{equation}
to represent the stationary distribution of a trajectory $\gamma\in\mathcal{M}$ in the forward process (starting from $\gamma(0)=I$ to $\gamma(T)=U$). A superscript ``tilde'' denotes the quantities related to the reverse process. By implementing time-reverse, such that $t\to\tilde{t}= T-t$ and $\gamma\to\tilde{\gamma}(\tilde{t})=\gamma(t)$ for $P(\gamma)$, we define the stationary distribution of a time-reverse trajectory $\tilde{\gamma}\in\mathcal{M}$ (starting from $\tilde{\gamma}(0)=U$ to $\tilde{\gamma}(T)=I$) as
\begin{equation}\label{4.8}
\tilde{P}(\tilde{\gamma})\equiv \tilde{P}(\tilde{\gamma},\tilde{T})=\frac{e^{-\eta C(\tilde{\gamma})}}{\tilde{Z}_a(\tilde{0})},
\end{equation}
where $\tilde{Z}_a(\tilde{0})$ refers to the partition function meeting the initial conditions of the reverse process. In the common thermodynamics' case, the stationary distribution of a forward trajectory $\vec{x}$ in a phase space \cite{70}
\begin{equation}\label{4.9}
P(\vec{x})=P(\vec{x},t)=\frac{e^{-\beta H(\vec{x})}}{Z(0)},
\end{equation}
and the integral is over the phase space. Similar to Eq. (\ref{4.8}) the counterpart of Eq. (\ref{4.9}), $\tilde{P}(\tilde{\vec{x}})$ represents the stationary distribution of a time-reverse trajectory $\tilde{\vec{x}}$ in the phase space.

$D(p(x)||q(y))$ represents the relative entropy between any two distributions $p(x)$ and $q(y)$
\begin{equation}\label{4.22}
D(p(x)||q(y))\equiv \int  p(x)\mathrm{log}\frac{p(x)}{q(y)}\mathrm{d}x\ge 0,
\end{equation}
which is equal to zero if and only if $p(x)=q(y),\forall x,y$. Such a quantity provides a measure of distinguishability and is a handy tool for quantifying time-asymmetry in thermodynamics.

\textit{Clarification}: our discussion on \textit{the second law of complexity} is an extension of that made by Brown and Susskind \cite{8}. However, there are two main differences between our discussion and theirs. The first difference is that the system $\mathcal{A}$ they discussed was a classical chaotic system. The Hamiltonian of the corresponding quantum system $\mathcal{Q}$ was time-independent. In contrast, our system $\mathcal{A}$ is a classical stochastic system, whose dual quantum system $\mathcal{Q}$ has a time-dependent Hamiltonian. The second difference is that we used different methods to study \textit{the second law of complexity}. In particular, we used the approach developed by Brock and Esposito \cite{28} \textit{the trajectory thermodynamics} to explore \textit{the second law of complexity} by analogy with their discussions on the second law of thermodynamics of nonequilibrium systems.

\subsubsection{Second law of thermodynamics and trajectory thermodynamics}\label{C421}
This section provides a brief review on \textit{the second law of thermodynamics} for nonequilibrium systems and \textit{the trajectory thermodynamics} \cite{28}. Note that the most common expression of \textit{the second law of thermodynamics} is known as the \textit{Clausius inequality}, that is,
\begin{equation}\label{4.10}
\beta  \langle Q\rangle\le \Delta S,
\end{equation}
where $Q$ is the heat absorbed by the system during a process and $\beta$ and $S$ represent the constant inverse temperature and the system's thermodynamic entropy, respectively. We define the free energy of the system as
\begin{equation}\label{4.11}
F\equiv U-\beta^{-1}S,
\end{equation}
where $U$ denotes the system's internal energy. By combining this definition with \textit{the first law of thermodynamics},
\begin{equation}\label{4.12}
\Delta U= W + Q,
\end{equation}
we obtain
\begin{equation}\label{4.13}
\langle W\rangle \ge \Delta F,
\end{equation}
which corresponds to the Kelvin-Planck statement of the second law of thermodynamics: \textit{it is impossible to extract energy from a sole heat bath and converse all that energy into work without introducing any other influence.} Equivalently, Eq. (\ref{4.13}) can be written as
\begin{equation}\label{4.14}
\langle \Delta S_{total}\rangle=\langle \Delta_iS\rangle\equiv\beta \langle W_{diss}\rangle\equiv\beta(\langle W\rangle-\Delta F)\ge 0,
\end{equation}
where $\langle\Delta S_{total}\rangle$ denotes the combined entropy change of the system and environment \cite{70} and $\langle W_{diss}\rangle\equiv\langle W\rangle-\Delta F$ is the average dissipated work for the forward process\footnote{Because $\langle W_{diss}\rangle$ is a physical measure quantifying the dissipation, Eq. (\ref{4.14}) is also a measure of dissipation.}. $\Delta_iS$ is defined as the cumulative entropy production along a trajectory \cite{28}, which is the time integration of the entropy production $\dot{S}_i\equiv \frac{\mathrm{d} S_i}{\mathrm{d}t}$. Therefore, the non-negativity of $\Delta_iS$ can be converted into the following form:
\begin{equation}\label{4.15}
\langle \dot{S}_i\rangle\ge 0
\end{equation}
which is one of the basic features of the thermodynamic second law \cite{28}. Eq. (\ref{4.13}) can also be derived from Eq. (\ref{1.2}) by directly applying the Jensen's inequality, that is, $\langle e^x\rangle\ge e^{\langle x\rangle}$. Thus, Eq. (\ref{1.2}) is closely related to \textit{the second law of thermodynamics}.

Next, we review the second piece of the ``puzzle,'' that is, \textit{the trajectory thermodynamics} \cite{28}. The cumulative entropy production along a forward trajectory $\vec{x}$ in phase space is defined as the log-ratio of the distributions for observing its trajectory in the forward and reverse processes.
\begin{equation}\label{4.16}
\Delta_i S(\vec{x})=\mathrm{log}\frac{P(\vec{x})}{\tilde{P}(\tilde{\vec{x}})}=\beta(W-\Delta F),
\end{equation}
where $W$ denotes the work done on the system in a forward experiment. Instead of the trajectories in the phase space, we treat the cumulative entropy production as a random variable because it encodes each trajectory. Consequently, the distribution of the cumulative entropy production is given by the path integral in phase space in combination with Eq. (\ref{4.9}):
\begin{equation}\label{4.17}
\begin{aligned}
P(\Delta_i S)&\equiv\int \delta (\Delta_i S-\Delta_i S(\vec{x}))P(\vec{x})\mathrm{d}\vec{x}\\
&=\mathrm{exp}(\Delta_i S)\int \delta (\Delta_i S-\Delta_i S(\vec{x}))\tilde{P}(\tilde{\vec{x}})\mathrm{d}\vec{x}\\
&=\mathrm{exp}(\Delta_i S)\int \delta \left(-\Delta_i S-\Delta_i \tilde{S}(\tilde{\vec{x}})\right)\tilde{P}(\tilde{\vec{x}})\mathrm{d}\tilde{\vec{x}}\\
&\equiv\mathrm{exp}(\Delta_i S)\tilde{P}(-\Delta_i S),
\end{aligned}
\end{equation}
and because $\tilde{\tilde{\vec{x}}}=\vec{x}$ and $\tilde{\tilde{P}}=P$, the cumulative entropy production along a reverse trajectory $\tilde{\vec{x}}$ is obtained by
\begin{equation}\label{4.18}
\Delta_i \tilde{S}(\tilde{\vec{x}})=\mathrm{log}\frac{\tilde{P}(\tilde{\vec{x}})}{P(\vec{x})}=-\Delta_i S(\vec{x}).
\end{equation}
Furthermore, because the Jacobian for the transformation to the time-reverse variables is equal to one \cite{28}, we can conclude from Eq. (\ref{4.17}) that
\begin{equation}\label{4.19}
\frac{P(\Delta_i S)}{\tilde{P}(-\Delta_i S)}=\mathrm{exp}(\Delta_i S),
\end{equation}
which is called detailed fluctuation theorem \cite{77}. Eq. (\ref{4.19}) has the corresponding statement \cite{28}: the probability of stochastic entropy's increase in the forward process is exponentially more probable than that of a corresponding decrease in the reverse process. We can rewrite the Eq. (\ref{4.19}) as
\begin{equation}\label{4.20}
\langle\mathrm{exp}(-\Delta_i S)\rangle=1
\end{equation}
by integrating $\Delta_i S$ in  Eq. (\ref{4.19}). Hence by directly applying Jensen's inequality to Eq. (\ref{4.20}), we obtain Eq. (\ref{4.14}). Finally, we note that the average cumulative entropy production can be given by:
\begin{equation}\label{4.21}
\langle \Delta_i S\rangle=\beta(\langle W\rangle-\Delta F)=D(P(\vec{x})||\tilde{P}(\tilde{\vec{x}}))\ge 0,
\end{equation}
where $D(P(\vec{x})||\tilde{P}(\tilde{\vec{x}}))$ represents the relative entropy between $P(\vec{x})$ and $\tilde{P}(\tilde{\vec{x}})$.

\subsubsection{Discussions on the second law of complexity}\label{C422}
\textit{The second law of complexity} was first conjectured in \cite{44} and developed in \cite{8,67}. This conjecture has two equivalent statements:
\begin{itemize}
\item[1.]
Conditioning on the complexity being less than maximal, it will most likely increase, both into the future and into the past (Statement 1).
\end{itemize}
\begin{itemize}
\item[2.]
Decreasing complexity is unstable (Statement 2).
\end{itemize}
These statements initially described the features of complexity growth for a chaotic auxiliary system, which is dual to a quantum system with a time-independent Hamiltonian. To avoid confusion, we stipulate that the first statement is called ``Statement 1,'' and the second statement is called ``Statement 2.'' We mainly focus on Statement 2. Analogous to the canvass in Section \ref{C421}, we extend the discussion on \textit{the second law of complexity} to the case in which system $\mathcal{A}$ is stochastic (i.e., corresponds to a quantum system with a time-dependent Hamiltonian). Notably, we argue that the complexity version of the Jarzynski identity and \textit{trajectory thermodynamics} provide a ``Kelvin-Planck-like'' statement and a new version of Statement 2 of \textit{the second law of complexity} for stochastic auxiliary systems.

The distribution for a forward trajectory $\gamma$ in $\mathcal{M}$ is presented as Eq. (\ref{4.7}). Moreover, the distribution for its reverse $\tilde{\gamma}$ is represented by Eq. (\ref{4.8}). By analogy with Eq. (\ref{4.16}) we introduce a new quantity similar to the cumulative entropy production as follows:
\begin{equation}\label{4.23}
\Delta_i C(\gamma)=\log{\frac{P(\gamma)}{\tilde{P}(\tilde{\gamma})}},
\end{equation}
and we refer to it as the \textit{cumulative complexity production} along the forward trajectory $\gamma$. Moreover, we replace $\Delta_i S$, work $W$, and free energy difference $\Delta F$ in Eq. (\ref{4.21}) with Eq. (\ref{4.23}), computational work $W_a$, and computational free energy difference $\Delta F_a$, respectively. We obtain
\begin{equation}\label{4.24}
\langle\Delta_i C\rangle=\eta\left(\langle W_a\rangle-\Delta F_a\right)=-\eta\left(\langle C\rangle+\Delta F_a\right)\ge 0.
\end{equation}
This inequality can be obtained by applying the Jensen's inequality to Eq. (\ref{3.32}). Thus, the complexity version of the Clausius inequality is obtained as follows:
\begin{equation}\label{4.25}
\langle C\rangle\le -\Delta F_a.
\end{equation}
Because Eq. (\ref{4.24}) and Eq. (\ref{4.25}) are similar to Eqs. (\ref{4.14}) and (\ref{4.13}), respectively, we conclude that Eqs. (\ref{4.24}) and (\ref{4.25}) are the mathematical expressions of \textit{the second law of complexity} for stochastic auxiliary systems. Eq. (\ref{4.15}) denotes the equivalent expression of Eq. (\ref{4.14}), which describes the increase of entropy for nonequilibrium systems. After making an analog with Eq. (\ref{4.14}), Eq. (\ref{4.24}) corresponds to a ``Kelvin-Planck-like'' statement of the second law of complexity and describes the increasing nature of complexity, which is the stochastic generalization of \textit{the second law of complexity}.

Let us consider $\Delta_i C(\gamma)$ as a random variable. Analogous to Eq. (\ref{4.17}), the resulting distribution for $\Delta_i C$ can be obtained by doing a path integral in $\mathcal{M}$:
\begin{equation}\label{4.26}
\begin{aligned}
P(\Delta_i C)&\equiv\int \delta (\Delta_i C-\Delta_i C(\gamma))P(\gamma)D\gamma\\
&=\mathrm{exp}(\Delta_i C)\int \delta (\Delta_i C-\Delta_i C(\gamma))\tilde{P}(\tilde{\gamma})D\gamma\\
&=\mathrm{exp}(\Delta_i C)\int \delta (\Delta_i C+\Delta_i \tilde{C}(\tilde{\gamma}))\tilde{P}(\tilde{\gamma})D\tilde{\gamma}\\
&\equiv\mathrm{exp}(\Delta_i C)\tilde{P}(-\Delta_i C)
\end{aligned}
\end{equation}
that can be rewritten as
\begin{equation}\label{4.27}
\frac{P(\Delta_i C)}{\tilde{P}(-\Delta_i C)}=\mathrm{exp}(\Delta_i C).
\end{equation}
Eq. (\ref{4.27}) is the complexity version of the detailed fluctuation theorem analogous to Eq. (\ref{4.19}) and the mathematical expression of the new version of Statement 2 for the stochastic system $\mathcal{A}$. We summarize the statement as follows: \textit{the possibility for an increase in stochastic complexity production is exponentially greater than that of a corresponding decrease}. The average cumulative complexity production is equal to zero if and only if $P(\gamma)=P(\tilde{\gamma})$, $\forall \gamma\in\mathcal{M}$. This reveals that the time-asymmetry only vanishes when all possible trajectories in $\mathcal{M}$ are reversible (i.e., the maintenance of time-reversal symmetry). However, this vanishing condition is extremely hard to satisfy because the probability of reversing even a small fraction of the trajectory in the space of unitary operators is negligible for a stochastic system $\mathcal{A}$, which has been already discussed in Section 9 of \cite{67}.

This section ends with the exploration of the probability to observe a complexity value that rises beyond the complexity upper bound obtained by applying the Jensen's inequality to Eq. (\ref{3.32}), that is, $-\Delta F_a$. Same as before, let us resolve this problem within thermodynamics and give a thermodynamic analog of the complexity. To experimentally obtain the average work, we measure the fluctuating work $W_i$ of a single trajectory in a specific carry out $i$ of the experiment in statistical mechanics \cite{69}. A protocol defines a family of the Hamiltonian $\{H(t)\}$ governing the system evolution from $t=0$ to $t=T$. The experiments are run by controlling a parameter, such as
\begin{equation}\label{4.29}
\Lambda(t) \equiv \Lambda_0 + (\Lambda_{T}-\Lambda_0)\times \frac{t}{T},\quad \mathrm{with} \quad 0\le t \le T,
\end{equation}
where $\Lambda=\{\Lambda_i\}$ is a set of external controlled parameters changing in time. One can consider a similar situation in the complexity context, because the trajectories in $\mathcal{M}$ are generated by time-dependent Hamiltonians and each trajectory corresponds to a specific value of complexity. In particular, we consider a similar protocol that defines a family of $\{H(t)\}$ and evolves the unitary operator\footnote{Note that we evolve a unitary operator using a time-dependent Hamiltonian instead by evolving a quantum state.} in a time interval $T$. Then, we repeat the experiment $n$ times and compute the complexity for each experiment of the quantum system $\mathcal{Q}$\footnote{We must reinitialize the system $\mathcal{Q}$ to the same initial state after each experiment.}. Taking the arithmetic mean of these values to build a complexity ensemble (i.e., $\Omega=\{C_i\}$), we can construct a probability distribution $P(C)$ for complexity in the limit $n\to\infty$. The average complexity is obtained as
\begin{equation}\label{4.30}
\langle C(T)\rangle=\underset{n\to \infty}{\mathrm{lim}}\frac{1}{n}\sum_{i=1}^{n}C_i(T)=\int P(C)C\mathrm{d}C.
\end{equation}
Notably, the considered Hamiltonians are time-dependent and the evolution of the system $\mathcal{A}$ follows a Markov process\footnote{If the system has a time-independent Hamiltonians (e.g.,  SYK model), the randomness in the distribution function $P(C)$ comes from the random couplings $\{J\}$. We leave this for our future study.}. Now, suppose several experiments with $C>-\Delta F_a+\zeta$ are included in the ensemble $\Omega$. In combination with Eq. (\ref{3.32}), the probability of their appearance is as follows:
\begin{equation}\label{4.31}
\begin{aligned}
p_v[C>-\Delta F_a+\zeta]&\equiv\int_{-\Delta F_a+\zeta}^{\infty} P(C)\mathrm{d}C\\
&\le \int_{-\Delta F_a+\zeta}^{\infty} \mathrm{exp}[\eta(C+\Delta F_a-\zeta)]P(C)\mathrm{d}C\\
&\le \mathrm{exp}[\eta(\Delta F_a-\zeta)]\int_{0}^{\infty} \mathrm{exp}(\eta C)P(C)\mathrm{d}C\\
&=\mathrm{exp}(-\eta\zeta),
\end{aligned}
\end{equation}
where $\zeta$ is an arbitrary positive number. Eq. (\ref{4.31}) shows a behavior similar to a thermodynamic case \cite{70} where the left tail of the distribution $P(C)$ becomes exponentially suppressed in the forbidden region $C>-\Delta F_a$. Consequently, it is hard to measure a complexity value that rises significantly more than the multiples of $\eta$ beyond $-\Delta F_a$, which can be considered as a phenomenon related to \textit{the second law of complexity}. Unlike nonequilibrium thermodynamics \cite{76}, the lower limit of the integral in Eq. (\ref{4.31}) is not negative infinity but zero because the complexity metric should be non-negative \cite{2}.

\subsection{Fluctuation-dissipation theorem and complexity}\label{C43}
Because we obtained a fluctuation theorem for the complexity in Eq. (\ref{4.27}), it is interesting to ask whether it is possible to relate the complexity fluctuation to the cumulative complexity production representing the dissipation of the system $\mathcal{A}$ by analogy with thermodynamic fluctuation-dissipation theorem. To answer this question, we propose a complexity version of the fluctuation-dissipation theorem in this section.

Let us start by discussing the situation in thermodynamics. Using Eq. (\ref{1.2}) and the nonequilibrium work to derive the equilibrium free energy difference, we obtain
\begin{equation}\label{4.32}
-\beta\Delta F=\mathrm{log}\langle \mathrm{exp}(-\beta W)\rangle.
\end{equation}
We can expand the series on the left-hand side of this equation to the second-order term.
\begin{equation}\label{4.33}
-\beta\Delta F=(-\beta)\langle W\rangle+\frac{1}{2!}(-\beta)^2[\langle W^2\rangle-\langle W\rangle^2],
\end{equation}
where the second cumulant is
\begin{equation}\label{4.34}
\sigma_S^2\equiv (\beta)^2[\langle W^2\rangle-\langle W\rangle^2]
\end{equation}
as the variance of entropy $S$. Notably, Eq. (\ref{4.34}) represents the fluctuation of the entropy $S$. The discussion on the entropy production can be alternatively phrased in terms of dissipation work (i.e., $W_{\mathrm{diss}}=\beta^{-1} \Delta_i S$). From the Callen-Welton theorem \cite{71},
\begin{equation}\label{4.35}
\mathrm{dissipation}\propto\mathrm{fluctuation},
\end{equation}
the fluctuation-dissipation theorem can be obtained by combining Eq. (\ref{4.33}) with Eq. (\ref{4.16}), namely
\begin{equation}\label{4.36}
2\langle \Delta_i S\rangle=\sigma_S^2.
\end{equation}
This fluctuation-dissipation theorem was studied in \cite{72} and has many potential applications, including the construction of some hydrodynamic approaches \cite{57}.

We now go back to the content of complexity. Eq. (\ref{3.32}) gives
\begin{equation}\label{4.37}
-\eta\Delta F_a=\mathrm{log}\langle \mathrm{exp}(\eta C)\rangle,
\end{equation}
where the average value of the exponential function on the right-hand side is obtained using the probability $P(C)$:
\begin{equation}\label{4.38}
\langle \mathrm{exp}(\eta C)\rangle=\int \mathrm{exp}(\eta C)P(C)\mathrm{d}C.
\end{equation}
We expand the left-hand side of Eq. (\ref{4.37}) into an infinite series:
\begin{equation}\label{4.39}
-\eta\Delta F_a=\eta\zeta_1(C)+\frac{(\eta)^2}{2!}\zeta_2(C,C^2)+\sum_{n=3}^{\infty}\frac{(\eta)^n}{n!}\zeta_n(C,C^2,...,C^n),
\end{equation}
where $\zeta_n$ here denotes $n$-order cumulant\footnote{Note the distinction between $\zeta_n$ and the positive number $\zeta$ in Section \ref{C42}.}. Among these cumulants, the first and second order terms are
\begin{equation}\label{4.40}
\zeta_1(C)=\langle C\rangle,\quad \zeta_2=\sigma_C^2=\langle C^2\rangle-\langle C\rangle^2,
\end{equation}
and the second-order cumulant denotes the variance (fluctuation) of the complexity. Recall that the state space is $\mathcal{S}=\{\gamma_0,\gamma_1,\cdots\cdots,\gamma_{N}\}$ that contains $(N+1)$ mutually independent random variables. If we take the limit $N\to\infty$, then $P(C)$ is approximately Gaussian for the central limit theorem.
\begin{equation}\label{4.41}
P(C)\approx\frac{1}{\sigma^2_C\sqrt{2\pi}}\mathrm{exp}\left(-\frac{1}{2}\frac{[C-\langle C\rangle]^2}{\sigma^2_C}\right).
\end{equation}
Moreover, in this case, we can expand Eq. (\ref{4.37}) only to the second-order terms. Then, Eqs. (\ref{4.39}) and (\ref{4.24}) imply
\begin{equation}\label{4.42}
2\langle \Delta_i C\rangle=\eta^2\sigma_C^2.
\end{equation}
This is the version of the fluctuation-dissipation theorem for complexity that connects the fluctuation of complexity with the dissipation of the auxiliary system $\mathcal{A}$ during the evolution.

Notably, Eq. (\ref{4.42}) essentially links the fluctuation of the trajectories (each trajectory corresponds to a specific complexity value) in $\mathcal{M}$ with the time-dependent perturbation applied to the quantum system $\mathcal{Q}$ because any trajectory in $\mathcal{M}$ is generated by the Hamiltonian $H(t)$ of the system $\mathcal{Q}$. This connection implies that Eq. (\ref{4.42}) may play a vital role in quantifying holographic fluctuations, which will be discussed in Section \ref{C44}.

\subsection{Remarks on holographic fluctuations and complexity}\label{C44}
The discussions in this section are inspired by the remarkable work of Chemissany and Osborne \cite{21}, who developed a method for identifying the relation between the fluctuation of the bulk geometry and the perturbation applied to the boundary quantum system via \textit{the principle of minimal complexity}. We argue that the obtained Jarzynski framework provides a potential tool for quantitatively investigating the holographic fluctuations. We divide this section into two subsections. First, we briefly review the settings and main contribution of \cite{21}. Second, we give a remark based on the Jarzynski framework obtained in the previous sections.

\subsubsection{Basic settings and construction of the bulk spacetime}\label{C441}
The boundary system is a $2$-local quantum system comprising $K$ distinct subsystems (qubits)\footnote{For simplicity, we only consider the system has $2$-local Hamiltonian. In principle, one can consider a $k$-local Hamiltonian for any $k$.}, which is initialized in a trivial reference state $\ket{\Omega}$. We use different numbers that form a point set $\{1,2\cdots,K\}$ to label different subsystems. A unitary operator $U$ is generated in a certain time interval $T$ that diagonalizes a Hamiltonian $H$ of the boundary system. The focus here is the evolution of the unitary operator from $I$ to $U$. It is equivalent to the evolution of the trajectory $\gamma$ of a fictitious particle with unit mass (of the system $\mathcal{A}$) moving on $\mathcal{M}$ from $I$ to $U$. The time interval $T$ forms another set $[0,T]$, and a topological space is appointed as the bulk spacetime, i.e., $(\mathcal{X},\mathcal{T})$, where $\mathcal{X}=\{1,2\cdots\cdots,K\}\times[0,T]$ and $\mathcal{T}$ is an undetermined topology denoting the causality of the bulk spacetime. The point set $\mathcal{X}$ corresponds to holographic spacetime with discrete boundary spatial coordinates $j\in\{1,2,\cdots\cdots,K\}$ and ``radial'' holographic time coordinates as $t\in[0,T]$. We can completely identify a bulk spacetime from the trajectories $\{\gamma\}$ via \textit{the principle of minimal complexity} by determining the topology $\mathcal{T}$\cite{21}.

The target unitary $U$ form for the boundary system is presented in Eq. (\ref{2.1}). This expression can be approximately replaced by a discrete quantum circuit $U\approx V=V_T\cdots V_2V_1$, where $V_t,t\in\{1,2,\cdots\cdots,T\}$ denote the gates acting on one or two qubits at a moment. Therefore, the set $\mathcal{X}$ becomes $\mathcal{X}=\{1,2\cdots\cdots,K\}\times\{1,2,\cdots\cdots,T\}$. This forms a simple graph in Fig. 4.
\begin{figure}[htbp]
\centering
\includegraphics[scale=0.75]{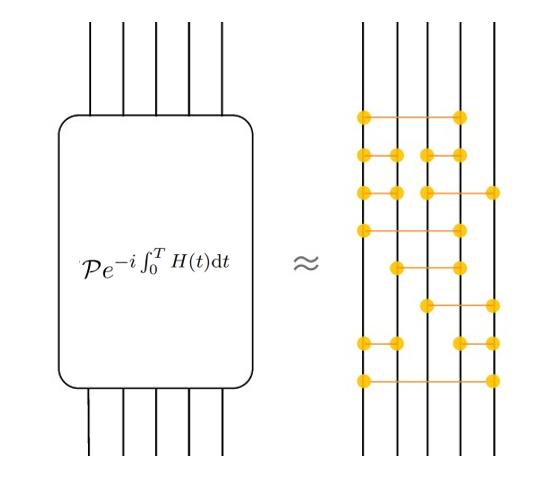} \label{P43}
\caption{Each line represents a qubit and edges connecting two lines refer to local gates.}
\end{figure}
We put an edge between the two vertices if a two-qubit gate acts nontrivially on a pair of qubits. To obtain the topology (causality) of the bulk spacetime, we first sample points from a Poisson distribution on set $\mathcal{X}$ with density $\rho$ to give a new finite set $\mathcal{Y}$. The causality relation on $\mathcal{Y}$ is then constructed by sending a detectable signal from a spacetime point $x=(i,s)$ to another point $y=(j,t)$ via a unitary process $\gamma$. We are allowed to interrupt the evolution of $\gamma$ by introducing arbitrary fast local interventions\footnote{Local unitary operations introduce these interventions \cite{21}.} at any holographic time $t=t_{w}$. Consequently, this method of building causal structures connecting with trajectory $\gamma$ gives us a topology for building the topological space $(\mathcal{X},\mathcal{T})$ regarded as the bulk spacetime.

\subsubsection{Holographic fluctuations and Jarzynski identity}\label{C442}
According to the above discussion, any geodesic $\gamma\in \mathcal{M}$ gives rise to the bulk spacetime. Therefore, the fluctuating trajectories in $\mathcal{M}$ (i.e., the trajectories with near-minimal complexity) can be interpreted as fluctuations in the bulk geometry considered as holographic fluctuations. To capture structures of holographic fluctuations, we only need to describe the structures of the fluctuating trajectories in $\mathcal{M}$ based on the three following premises:
\begin{itemize}
\item[1.]
The complexity, $C(\gamma)$, is sensitive to the applied 2-local interactions (quantum gates) between an arbitrary pair of qubits, but not to a particular pair of qubits to which the unitary gate is applied \cite{21}.
\end{itemize}
\begin{itemize}
\item[2.]
A complexity functional determines a geodesic in $\mathcal{M}$ similar to an action functional specifies a geodesic in classical mechanics.
\end{itemize}
\begin{itemize}
\item[3.]
Any trajectory in $\mathcal{M}$ arises from the boundary system via the $\mathcal{Q}$-$\mathcal{A}$ correspondence; thus, perturbing the boundary system by inserting quantum gates is equivalent to perturbing the trajectory $\gamma$ in $\mathcal{M}$.
\end{itemize}
Fig. 5 depicts the structure of the holographic fluctuations summarized as follows: the trajectories are equal to $\gamma(t)$ for all $t$, except at one moment $t=t_w$ when a local unitary gate\footnote{One can regard this gate as the arbitrary fast local intervention.} is applied to an arbitrary pair of qubits $i$ and $j$, followed immediately by its inverse gate \cite{21}. Because the applied gate generates an instantaneous interaction between qubits $i$ and $j$ and the inverse gate cancels the interaction effect, a ``wormhole'' is created between two points $(i,t_w)$ and $(j,t_w)$ in the dual bulk spacetime that immediately ``evaporates''. In \cite{21}, Eq. (\ref{3.9}) was introduced to model these fluctuations. Recall that the complexity has a quadratic action form in Eq. (\ref{3.9}); hence, the fluctuating trajectories in $\mathcal{M}$ can be understood as the stochastic trajectories of the Brownian motions in $\mathcal{M}$ and are the solutions of Eq. (\ref{2.22}) invoked as a toy model of the black hole in \cite{46}. In summary, the bulk geometry modeled by Eq. (\ref{3.9}) constitutes a spacetime where ``wormholes'' are fluctuating in and out of existence between all pairs of spacetime points \cite{21}.
\begin{figure}[htbp]
\centering
\includegraphics[scale=0.75]{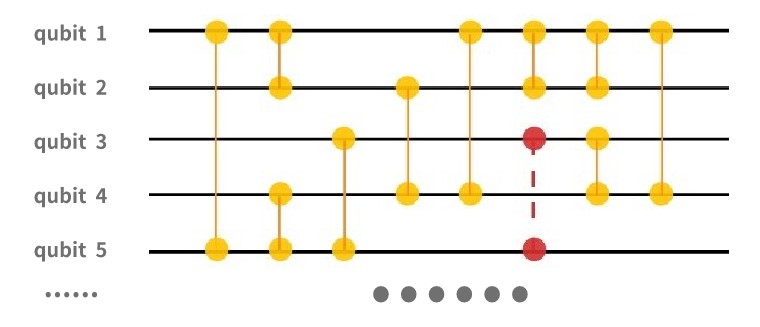} \label{P42}
\caption{This can be seen as the structure of holographic fluctuation which is composed of qubits (lines) and gates (connections between lines). The red part represents a fluctuation, that is, in a time-slice, a $2$-local gate is applied to an arbitrary pair of qubits followed by its inverse.}
\end{figure}

Now, we make a remark on the holographic fluctuations from the perspective of the obtained Jarzynski framework. Applying the results we obtained in the previous sections helps us obtain a better quantitative realization of the holographic fluctuations and several clues strengthen our confidence about that. First, the holographic fluctuation follows a stochastic process in $\mathcal{M}$, which means that we can introduce a partition function Eq. (\ref{3.9}) using the path integral approach to model its structure. Because we have the partition function, the Jarzynski identity Eq. (\ref{3.32}) can be obtained to further give a framework that provides us with a version of the fluctuation theorem for complexity, Eq. (\ref{4.27}), which describes the complexity fluctuations. Second, the complexity fluctuations can equivalently describe the fluctuations of trajectories in $\mathcal{M}$ based on the second premise, and the fluctuating trajectories can give rise to the bulk spacetime. Therefore, the fluctuation theorem describes the fluctuations of complexity and bulk geometry. Third, the fluctuation-dissipation theorem connects the average cumulative complexity production with fluctuations of complexity, which is usually used to study the response of a system to external influences. Thus, Eq. (\ref{4.42}) may be applicable to detect the response of the bulk geometry to some perturbations applied to the boundary quantum system. In particular, this equality can be used to quantitatively measure the fluctuations of the complexity by capturing the information on the average cumulative complexity production.

\section{Example: transverse field Ising model}\label{C5}
The obtained Jarzynski framework gives us few interesting conclusions that need testing. Camilo and Teixeira \cite{73} studied the complexity of the transverse field Ising model (TFIM). We follow their steps to numerically test two of our main proposals, Eqs. (\ref{4.24}) and (\ref{4.25}). For simplicity, we only consider two phases with the ferromagnetic order along the $z$ direction (FMZ) and the paramagnetic phase (PM). We do not focus on the detailed derivation here but will only present a brief review of the derivation with minimal efforts. One can refer to \cite{71} for a detailed derivation.

\subsection{Model settings}\label{C51}
The TFIM is determined as follows by the time-dependent Hamiltonian
\begin{equation}\label{5.1}
H(t)=-J\sum_{j=1}^N \sigma_j^3\sigma_{j+1}^3-g(t)\sum_{j=1}^N\sigma_j^1,
\end{equation}
where $J$ denotes the definite numbers representing couplings and $\sigma_j^3$ and $\sigma_j^1$ are the Pauli matrices acting on the $j$th lattice site. $g(t)=g_0+g_1(t)\mathrm{cos}(\xi t )$ is the transverse field denoting the perturbation comprising a constant $g_0$ and monochromatic driving term with frequency $\xi$. Assuming that the system is a closed lattice with periodical boundaries $\sigma_{N+1}^{\alpha}=\sigma_{1}^{\alpha}$, restricting $N$ to be even and applying the Fourier transformation, the Hamiltonian can be rewritten as follows in terms of Jordan-Wigner fermions $c_q=\frac{\mathrm{e}^{i\frac{\pi}{4}}}{\sqrt{N}}\sum_{q\in B}c_k\mathrm{e}^{ikq}$ as $H(t)=\sum_{k>0}H_k(t)$:
\begin{equation}\label{5.2}
H_k(t)=[2g(t)-\omega_k](c_k^{\dagger}c_k+c_{-k}^{\dagger}c_{-k})+\Delta_k(c_k^{\dagger}c_{-k}^{\dagger}+c_{-k}c_k)-\omega_k,
\end{equation}
where $B=\{\pm \frac{\pi}{N},\pm \frac{3\pi}{N},\pm \frac{5\pi}{N},\cdots\cdots,\pm \frac{(N-1)\pi}{N}\}$ denotes the Brillouin zone, $\omega_k=2J\mathrm{cos}k$, $\Delta_k=2J\mathrm{sin}k$, and the trivial contribution $-2Ng(t)$ is neglected. Eq. (\ref{5.2}) is called the Bogoliubov-de Gennes (BdG) Hamiltonian which conserves momentum and parity; the latter implements the $\mathbb{Z}_2$ symmetry resulting in a decomposition of the Hilbert space into a direct sum of Neveu-Schwarz (NS) sectors. The system evolution dynamically obeys the Schr\"{o}dinger's equation. The dynamics is confined to the two-level Nambu subspace spanned by $\{\ket{0_{-k}0_k},\ket{1_{-k}1_k}\}$. The system state at any time $t$ will acquire the following form:
\begin{equation}\label{5.3}
\ket{\Psi(t)}=\otimes_{k> 0}[u_k(t)\ket{1_{-k}1_k}+v_k(t)\ket{0_{-k}0_k}],
\end{equation}
where the coefficients follow the Schr\"{o}dinger equation, and the spinor is denoted by the symbol $\Psi_k(t)\equiv [u_k(t)$ $ v_k(t)]^{\mathrm{T}}$.

Imagine that the system is initialized in state $\ket{\Omega}=\otimes_{k> 0}\ket{0_{-k}0_k}$ at $t=0$ and evolves during a time interval $T$ to a target state $\ket{\Psi (T)}=U\ket{\Omega}$ through a specific unitary operator $U=\gamma(T)=\otimes_{k> 0}U_k$, where $U_k$ represents the $k$th momentum sector of $U$. The boundary conditions $\gamma_k(0)=I$ and $\gamma_k(T)=U_k$ are fixed. The application of the Bogoliubov transformation suggests that the complexity metric for each momentum sector is presented as follows in terms of Hopf coordinates $(\phi_1,\phi_2,\omega)$
\begin{equation}\label{5.4}
\mathrm{d}s^2|_k=\mathrm{d}\omega^2+\mathrm{cos}^2\omega \mathrm{d}\phi_1^2+\mathrm{sin}^2\omega \mathrm{d}\phi_2^2,
\end{equation}
where $\phi_1$ and $\phi_2$ correspond to two phases and
\begin{equation}\label{5.5}
\omega(t)=\frac{t}{T}\times\left |{\mathrm{arcsin}\left (\frac{\Delta_k\theta^{(l)}}{\epsilon_{(k,l)}}\mathrm{sin}(\epsilon_{(k,l)}t)\right )}\right |, \quad t\in[0,T]
\end{equation}
denotes the linear profile, where $l\in \mathbb{Z}$. The anisotropic parameter \cite{74,75} and the eigenvalues of BdG Hamiltonian are represented by
\begin{equation}\label{5.6}
\epsilon_{(k,l)}=\sqrt{(\delta g_0^{(l)}-\omega_k)^2+(\Delta_k\theta^{(l)})^2},\quad \theta^{(l)}= (-1)^l\mathcal{J}_l\left (\frac{4g_1}{\eta}\right),
\end{equation}
respectively. They are obtained from the Bogoliubov transformation and the high-frequency driving approximation \cite{73}. Here $\mathcal{J}_l(x)$ represents the Bessel functions and $\delta g_0^{(l)}\equiv g_0-l\xi/4$ is called the detuning parameter. After summing over all $k$ for Eq. (\ref{5.4}), from Eq. (\ref{2.18}), the complexity is derived  in the following form:
\begin{equation}\label{5.7}
\begin{aligned}
C(t)&=\underset{\gamma}{\mathrm{inf}}\sum_{k> 0} C_k(t)\\
&\equiv\frac{1}{2}\sum_{k> 0}\left |{\mathrm{arcsin}\left (\frac{\Delta_k\theta^{(l)}}{\epsilon_{(k,l)}}\mathrm{sin}(\epsilon_{(k,l)}t)\right )}\right |^2,
\end{aligned}
\end{equation}
where $C_k$ is defined as the complexity of the $k$th momentum sector solely.

A numerical simulation is performed after the parameters $N$, $l$, $J$ $g_l$, and $\delta g_0^{(l)}$ and the $\eta$ value ($\eta=1$) are set. We use time-average instead of ensemble-average in the simulation because the motion on $\mathcal{M}$ satisfies ergodicity. Therefore, we take the limit $T\to \infty$ and precisely replace the ensemble-average with the time-average:
\begin{equation}\label{5.8}
\left\langle Q_a \right\rangle_{\mathrm{ensemble}}=\left\langle Q_a(T) \right\rangle_{\mathrm{time}}\equiv \underset{T\to \infty}{\mathrm{lim}}\frac{1}{T}\int_0^TQ_a(t)\mathrm{d}t,
\end{equation}
where $Q_a$ represents two related quantities, that is, complexity $C$ and the cumulative complexity production $\Delta_i C$.

\subsection{Numerical results}\label{C52}
Two features of the numerical simulation support our previous analytical results:
\begin{itemize}
\item[1.]
The computational free energy difference provides an average complexity upper bound, which is not violated. This supports Eq. (\ref{4.25}).
\end{itemize}
\begin{itemize}
\item[2.]
The non-negative average cumulative complexity production supports Eq. (\ref{4.24}), which is the mathematical expression corresponding to the ``Kelvin-Planck-like'' statement of \textit{the second law of complexity}.
\end{itemize}

The Hamiltonian Eq. (\ref{5.1}) corresponds to various regimes according to the different values of the detuning parameter $\delta g_0^{(l)}$. The phases can be changed from the FMZ phase through a quantum critical point (QCP) to the PM phase by varying $\delta g_0^{(l)}$ from $0$ to $J$ and to $2J$ \cite{73}. For simplicity, we do not consider the critical behavior of the QCP herein and use Eq. (\ref{5.8}) to calculate the time-averaged values of the complexity-related quantities for the FMZ and PM phases.
\begin{figure}[htbp]\label{P52}
\centering
\subfigure[Evolution of Ferromagnetic Phase]{
\includegraphics[scale=0.3]{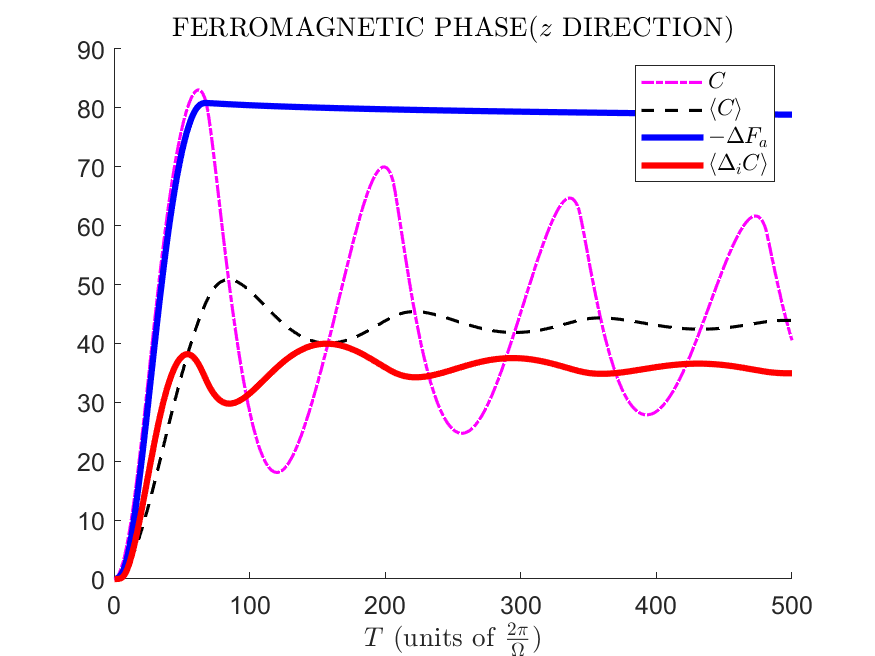} \label{P521}
}
\quad
\subfigure[Evolution of Paramagnetic Phase]{
\includegraphics[scale=0.3]{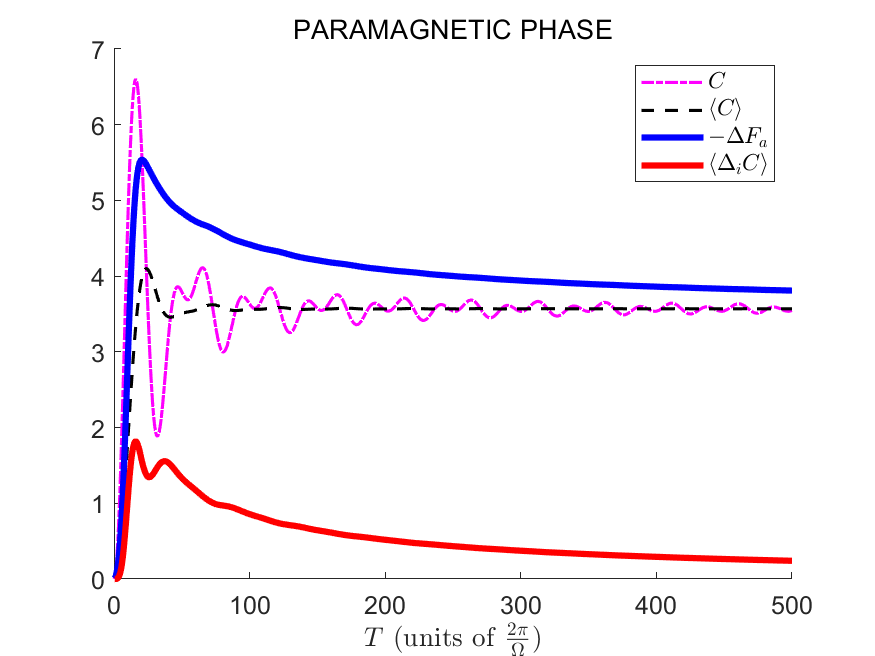} \label{P522}
}
\caption{Here $C$, $\langle C\rangle$, $\Delta F_a$ and $\langle \Delta_i C\rangle$ represent complexity, average complexity, the computational free energy difference and the average cumulative complexity production, respectively, and the parameters are set to $\eta=1$, $N=1000$, $l=2$, $J=0.01\Omega$, $g_1=\Omega$, and $\delta g_0^{(l)}=\{0, 2J\}$ that correspond to the ferromagnetic phase along the $z$ direction and the paramagnetic phase, respectively.}
\end{figure}
Recall that $\delta g_0^{(l)}=\{0,2J\}$ correspond to the FMZ and PM phases. We plot the complexity $C$, time-averaged complexity $\langle C\rangle$, computational free energy difference $-\Delta F_a$, and average cumulative complexity production $\langle \Delta_i C\rangle$, as a function of time interval $T$ in Figs. \ref{P521} and \ref{P522}.

The FMZ and PM phases present an approximately linear growth initially but subsequently show distinct behaviors. Because the FMZ phase is susceptible to the time-dependent transverse field, its complexity finds it hard to maintain stability and violently fluctuates around a certain value for a long time interval (Fig. \ref{P521}). In contrast, the complexity of the PM phase will remain stable around a certain value with time interval (Fig. \ref{P522}). The average cumulative complexity production of the PM phase gradually becomes more negligible and eventually turns to zero. Physically, this comes from the disordered character of the PM phase: ``non-local operations are required to create order in a state of the PM phase, but local operations would maintain disorder of such a state. Consequently, the influence of the transverse field is suppressed to prevent the system from creating non-local gates to order the system when $g_0$ is large \cite{73}.'' However, for the FMZ phase, even though the average cumulative complexity production will gradually decrease, it will not drop to zero for a long period. The computational free energy differences are obtained by applying the Jarzynski identity corresponding to the changes for the FMZ and PM phases shown in Figs. \ref{P521} and \ref{P522}, respectively.

We have simulated the dissipation for the FMZ (Fig. \ref{P521}) and PM (Fig.\ref{P522}) phases and obtained a fluctuation-dissipation for the complexity. Therefore, in Eq. (\ref{4.42}), we can make some further discussions about the dissipative behaviors of the two phases and their relation with the holographic fluctuations. First, let us discuss the evolution of the average cumulative complexity production (dissipative behaviors) for the large $T$ regime, where $P(C)$ is simply Gaussian because of the central limit theorem. Fig. \ref{P521} depicts that for the FMZ phase, no steady state can be found in a short period (the complexity violently fluctuates) and its average cumulative complexity production always takes large values. Meanwhile, the average cumulative complexity production of the PM phase (plotted in Fig. \ref{P522}) shows a downward trend and tends to be zero for $T\to\infty$, indicating that the dissipation vanishes for large $T$. Physically, this means that the quantum system $\mathcal{Q}$ reaches its average complexity upper bound (i.e., $\left\langle C\right\rangle\to -\Delta F_a$), such that no resource can be extracted from the system $\mathcal{Q}$ due to the breaking of the time-asymmetry for any possible trajectory in $\mathcal{M}$ \cite{28}, and reversibility holds for all possible trajectories. Additionally, since $\left\langle \Delta_i C\right\rangle=0$ can only be obtained when $T\to\infty$, we assume $T\to\infty$ as a complexity quasi-static limit analogous to thermodynamics. By taking this limit, Eq. (\ref{4.25}) takes the equal sign and the average complexity of system $\mathcal{Q}$ reaches its upper bound\footnote{Brown and Susskind called this ``complexity equilibrium'' \cite{8}. However, because ``equilibrium'' is usually used to describe macroscopic quantities in thermodynamics, to avoid confusion, we do not use this word in this study.}. As mentioned in Section \ref{C442}, the complexity fluctuations can be regarded as bulk geometry fluctuations and the fluctuation-dissipation theorem states that $\langle \Delta_i C\rangle\sim \sigma^2_C$; hence, theoretically, we can construct a bulk spacetime from a topological space and simulate its fluctuation using Eq. (\ref{4.42}). In particular, let the TFIM be our boundary quantum system $\mathcal{Q}$ with a time-dependent perturbation (transverse field). Note that the transverse field in the system $\mathcal{Q}$ causes the geodesics in $\mathcal{M}$ to fluctuate. We can then utilize the method of \cite{21} to construct a dual topological space from the TFIM as our bulk spacetime. The geometric structures of the bulk spacetime are changed because the transverse field affects the complexity to vary. We leave the exploration of this part to our future work.

We end this section with a remark. In the sense of average, Eq. (\ref{5.8}) may not be sufficiently accurate to describe behaviors of small $T$ regime  because it holds strictly only when $T\to\infty$. Therefore, performing the time average might not be the best approach to run simulations. In comparison, employing the Metropolis algorithm over Monte Carlo sweeps may be more practical in analogy with the cases of performing ensemble average, which has already been used to model a similar scenario of the common Jarzynski identity \cite{76}.

\section{Conclusions and Outlooks}\label{C6}
This study is motivated by the Nielsen's complexity geometry and the elegant proof of the Jarzynski identity done by Hummer and Szabo. We introduced the path integral in the context of the complexity geometry and used it to derive a complexity version of the Jarzynski identity. In addition, we made remarks on different complexity-related topics based on the obtained identity. The first remark is that $\mathrm{exp}(-\eta F_a(T))=\left\langle \mathrm{exp}(-\eta R(T))\right\rangle=\left\langle \mathrm{exp}\left[-\eta (C_{\mathrm{max}}-C(T))\right]\right\rangle$ provides us a new evidence of the existence of a well-defined resource theory of uncomplexity \cite{8,87}. The second and most crucial remark is an extension of the proposal made by Brown and Susskind, that is, \textit{the second law of complexity} \cite{8}. Our focus was slightly different from theirs such that the quantum system $\mathcal{Q}$ we considered is governed by a time-dependent Hamiltonian that forms a classical auxiliary system $\mathcal{A}$ with stochastic features. However, Brown and Susskind considered a quantum system with a time-independent Hamiltonian forming a chaotic auxiliary system. We extended their original second law to the case involving a stochastic auxiliary system. Based on \textit{the trajectory thermodynamics}, we argued that the complexity version of the Jarzynski identity provides two mathematical expressions of \textit{the second law of complexity}. Third, we derived a fluctuation-dissipation theorem for complexity by analogy with thermodynamics, which links the fluctuation of complexity to a crucial quantity, namely the average cumulative complexity production. The last remark is on holographic fluctuations. Because any geodesic in the space of unitary operators encodes a bulk spacetime with an extra dimension via \textit{the principle of minimal complexity} \cite{21} and any geodesic in the space of unitary operators corresponds to a value of complexity, the complexity fluctuations can play the role of bulk geometry fluctuations. Furthermore, our framework connects with the complexity fluctuations, therefore, our results can provide us with a new perspective on the exploration of holographic fluctuations by applying the complexity version of the Jarzynski framework.

We only touched some aspects of these issues, and extensive topics are waiting to be tackled. Several of them are presented below:
\begin{itemize}
\item[$\bullet$]
To explore the holographic fluctuations, one must sample points from the Poisson distribution on point set $\mathcal{X}$, which is a discrete process. Consequently, integrals in $\mathcal{M}$ are hard to solve. It is significant to ask if there is a proper continuum limit. Moreover, taking the continuum limit, the resulting bulk spacetime for CFTs should then converge to AdS \cite{21}.
\end{itemize}
\begin{itemize}
\item[$\bullet$]
In our discussions, the considered boundary system is a normal quantum system comprising qubits but not a standard quantum field theory. The path integral complexity \cite{78, 79} should be a candidate for generalizing our formalism to the quantum field theory. Choosing a suitable definition of quantum complexity would facilitate directly linking the quantum computational complexity with holographic complexity \cite{80}. This generalization may provide us with deeper insights into the AdS/CFT duality, e.g., for \textit{Complexity=Action} \cite{19,20} and \textit{Complexity=Volume} conjectures \cite{16,17,18}.
\end{itemize}
\begin{itemize}
\item[$\bullet$]
We chose the TFIM as an example. One would like to know if our results are applicable for other models, such as the SYK model \cite{37,38,39,40}, or if we can directly simulate \textit{Quantum Brownian Circuit}. The \textit{Quantum Brownian Circuit} is quite complicated, and a quantum simulation might be needed. As a reference, \cite{81} recently proposed a quantum simulation for calculating the Jarzynski identity.
\end{itemize}

\acknowledgments
We would like to thank  Peng Cheng, Shao-Feng Wu and Yu-qi Lei for helpful discussions. This work is partly supported by NSFC (No.11875184).

\appendix

\section{Fokker-Planck equations}
The time-dependent operators $L_t$ in Eqs. (\ref{2.23}) and (\ref{3.12}) are Fokker-Planck operators. Hence, understanding the derivation of the Fokker-Planck equations is helpful \cite{82}. We first review the common derivation of the Fokker-Planck equation and generalize it to the cases in a curved space equipped with a non-Euclidean metric \cite{83}. The latter can be directly used in Eq. (\ref{2.23}).

Let us consider the stochastic differential equations (SDEs)
\begin{equation}\label{A1}
\mathrm{d}x(t)=f(x,t)\mathrm{d}t+g(x,t)\mathrm{d}B(t),
\end{equation}
where $x$ denotes a stochastic variable, $f$ and $g$ are the functions of $x$, and $\mathrm{d}B(t)$ denotes the independent Brownian motion with unit variance per unit time. We introduce an arbitrary function $h(x)$ and use the Ito's rule to derive the Fokker-Planck equation:
\begin{equation}\label{A2}
\mathrm{d}h(x)=\left(\frac{\mathrm{d}h}{\mathrm{d}x}\right)f(x,t)\mathrm{d}t+\left(\frac{\mathrm{d}^2h}{\mathrm{d}x^2}\right)\frac{g^2(x,t)}{2}\mathrm{d}t+\left(\frac{\mathrm{d}h}{\mathrm{d}x}\right)g(x,t)dB(t).
\end{equation}
If we take averages on both sides, we immediately obtain:
\begin{equation}\label{A3}
\begin{aligned}
\frac{\mathrm{d}\langle h(x)\rangle}{\mathrm{d}t}&=\left\langle f(x,t)\left(\frac{\mathrm{d}h}{\mathrm{d}x}\right)\right\rangle+\left\langle \frac{g^2(x,t)}{2}\left(\frac{\mathrm{d}^2h}{\mathrm{d}x^2}\right)\right\rangle\\
&=\int_{-\infty}^{\infty}\left[f(x,t)\left(\frac{\mathrm{d}h}{\mathrm{d}x}\right)+\frac{g^2(x,t)}{2}\left(\frac{\mathrm{d}^2h}{\mathrm{d}x^2}\right)\right]P(x,t)\mathrm{d}x,
\end{aligned}
\end{equation}
where $P(x,t)$ represents the distribution function satisfying $P(x\to\pm\infty,t)=0$ and $\int_{-\infty}^{\infty}P(x,t)\mathrm{d}x=1$. Using the part-by-part integration, and $\langle h(x)\rangle=\int_{-\infty}^{\infty}P(x,t)h(x)\mathrm{d}x$, we obtain
\begin{equation}\label{A4}
\int_{-\infty}^{\infty}h(x)\frac{\partial P(x,t)}{\partial t}\mathrm{d}x=\int_{-\infty}^{\infty}h(x)\left\{-\frac{\partial}{\partial x}\left[ f(x,t)P(x,t)\right]+\frac{1}{2}\frac{\partial^2}{\partial x^2}\left[g^2(x,t)P(x,t)\right]\right\}\mathrm{d}x,
\end{equation}
where $h(x)$ is independent of $t$. This equality can be transformed into the Fokker-Planck equation, i.e.
\begin{equation}\label{A5}
\frac{\partial P(x,t)}{\partial t}=L_tP(x,t)=-\nabla\cdot\left[ f(x,t)P(x,t)\right]+\frac{1}{2}\nabla^2\left[g^2(x,t)P(x,t)\right],
\end{equation}
where the time-dependent operator $L_t$ refers to the Fokker-Planck operator and $\nabla$ represents the Nabla operator. Eq. (\ref{2.23}) is the Fokker-Planck equation for a vector Ito stochastic equation, in which $x\to\vec{x}$, $f\to \vec{f}$ and $g(x,t)$ become a matrix with the same dimensions of $\vec{x}$.

In the context of the complexity geometry, the configuration space is the group manifold $\mathcal{M}$ equipped with a non-Euclidean metric, Eq. (\ref{2.15}). Thus, we must introduce some modifications to derive the Fokker-Planck equation governing the time evolution of distributions on a Riemannian manifold. The first modification denotes the volume element
\begin{equation}\label{A6}
\mathrm{d}\vec{x}\to[\mathrm{d}\gamma],
\end{equation}
where $[\mathrm{d}\gamma]$ denotes the Haar measure, namely Eq. (\ref{3.1}), which makes the volume element independent of the choice of coordinate systems. The gradient, divergence, and Laplacian (or Laplacian-Beltrami operator \cite{83}) are modified accordingly.
\begin{equation}\label{A7}
\nabla h\to\mathrm{grad}(h)_M\equiv G_{MN}\partial^N h,
\end{equation}
\begin{equation}\label{A8}
\nabla\cdot\vec{f}\to\mathrm{div}\left(\vec{f}\right)\equiv\partial_M f^M,
\end{equation}
where $f^M$ denotes the component of vector $\vec{f}\in\mathcal{M}$ and
\begin{equation}\label{A9}
\nabla^2h\to\mathrm{div}\left(\mathrm{grad}(h)\right)\equiv|G_{MN}(\gamma)|^{-\frac{1}{2}}\partial_M\left(|G_{MN}(\gamma)|^{\frac{1}{2}}G^{MN}\partial_N h\right).
\end{equation}
$h$ represents an arbitrary function. Applying these modifications to the Fokker-Planck equation yields
\begin{equation}\label{A10}
\frac{\partial P(\gamma,t)}{\partial t}=L_tP(\gamma,t)=-\mathrm{div}\left[\vec{f}(\gamma,t)P(\gamma,t)\right]+\frac{1}{2}\mathrm{div}\left\{\mathrm{grad}\left[g^2(\gamma,t)P(\gamma,t)\right]\right\},
\end{equation}
as the Fokker-Planck equation for the group manifold $\mathcal{M}$, where $\vec{f}$ is the vector in $\mathcal{M}$. We generally regard Eq. (\ref{A10}) as the general form of the Fokker-Planck equation governing the time evolution of distributions in any curved space by considering $\gamma$ as a vector in that space.

We provide a special example, i.e. Eq. (\ref{2.22}), to obtain a better understanding of the Fokker-Planck equation. Note that $\frac{i}{\sqrt{8K(K-1)}}\sum_{j<k}\sum_{\alpha_j,\alpha_k=0}^{3}\sigma_j^{\alpha_j}\otimes\sigma_k^{\alpha_k}$ is independent of $\gamma$; thus, we use $g=g(\gamma,t)$ to represent it. Recall that $g$ is a generalized Pauli matrix that is an anti-Hermitian operator. Hence, the square of $g$ is given as follows:
\begin{equation}\label{A11}
g^2=\left(\frac{i}{\sqrt{8K(K-1)}}\sum_{j<k}\sum_{\alpha_j,\alpha_k=0}^{3}\sigma_j^{\alpha_j}\otimes\sigma_k^{\alpha_k}\right)^{\dagger}\left(\frac{i}{\sqrt{8K(K-1)}}\sum_{j<k}\sum_{\alpha_j,\alpha_k=0}^{3}\sigma_j^{\alpha_j}\otimes\sigma_j^{\alpha_k}\right).
\end{equation}
We notice that $\vec{f}(\gamma,t)=-\frac{1}{2}\gamma$. Substituting $\vec{f}$ and Eq. (\ref{A11}) into Eq. (\ref{A10}), we finally obtain the Fokker-Planck equation governing the time evolution of the distributions of the \textit{Quantum Brownian Circuit} in $\mathcal{M}$.

\section{Path integral in $\mathcal{M}$}\label{F2}
A path integral measure is required to do path integrals in $\mathcal{M}$ which is a curved manifold equipped with the complexity metric. This measure generally contains an additional curvature modification on its exponent because the vector operation of the two points on $\mathcal{M}$ is involved in the derivation of the path integral. However, because the additional term cannot be included in the measure, this takes the form of a partition function varying from Eq. (\ref{3.9}). \cite{52} provided a method for doing the path integral in a curved space without any additional curvature modification on the exponent. \cite{52} introduced a factor that is contained in the measure. Therefore, we briefly introduce this method here.

For consistency, we set $\ket{\gamma,t}$ as the eigenstates of the position operators denoted by symbol $\hat{X}^M$.
\begin{equation}\label{B1}
\hat{X}^M\ket{\gamma,t}=X^M(t)\ket{\gamma,t}.
\end{equation}
We consider that a particle in the system $\mathcal{A}$ evolves from $\gamma_1\in \mathcal{M}$ at $t=t_1$ to $\gamma_2\in \mathcal{M}$ at $t=t_2$. Before deriving a more general case, we first assume that if $\mathcal{M}$ is flat, the complexity metric reduces to the standard inner-product metric. We also suppose that a source term $J(t)$ is added to contribute to the complexity. The complexity $C$ then becomes $C_J$.
\begin{equation}\label{B2}
C_J(\gamma,\gamma_*)=C(\gamma)+\int_{t_1}^{t_2}J_M(t)\left(X^M(t)-X^M_*(t)\right)\mathrm{d}t
\end{equation}
by introducing a fixed point $\gamma_*\in M$, where $\left(X^M(t)-X^M_*(t)\right)$ represents the tangent vector to the geodesic from $\gamma$ to $\gamma_*$. To generalize Eq. (\ref{B2}), we replace this vector with the geodesic interval $\lambda(\gamma_*;\gamma)$ \cite{55,56}. By definition, it is presented in the following form:
\begin{equation}\label{B3}
\left(X^M_*(t)-X^M(t)\right)\to\lambda(\gamma_*;\gamma)\equiv \frac{1}{2}D^2(\gamma_*;\gamma),
\end{equation}
where $D(\gamma_*;\gamma)$ is the relative geodesic length between points $\gamma$ and $\gamma_*$; thus, the tangent vector to the geodesic at $\gamma_*$ refers to
\begin{equation}\label{B4}
\lambda^M(\gamma_*;\gamma)=G^{MN}(\gamma_*)\frac{\delta}{\delta X^N_*}\lambda(\gamma_*;\gamma).
\end{equation}
Because the source $J(t)$ transforms like a covariant vector at $\gamma_*$ independent of $\gamma$, Eq. (\ref{B2}) is changed to
\begin{equation}\label{B5}
C_J=C(\gamma)-\int_{t_1}^{t_2}J_M(t)\lambda^M(\gamma_*;\gamma)\mathrm{d}t.
\end{equation}

Moreover, the Schwinger action principle \cite{83,84} states that
\begin{equation}\label{B6}
\delta K_a(\gamma_2,t_2;\gamma_1,t_1)[J]=i\left\langle \gamma_2,t_2|\delta C_J|\gamma_1,t_1\right\rangle[J]=0,
\end{equation}
from which the equation of motion may be inferred as
\begin{equation}\label{B7}
\frac{\delta C}{\delta \lambda^M}(J)=J_M.
\end{equation}
We follow by expanding $K_a(\gamma_2,t_2;\gamma_1,t_1)[J]$ in the Taylor series about $J_M=0$ as
\begin{equation}\label{B8}
\begin{aligned}
K_a(\gamma_2,t_2;\gamma_1,t_1)[J]&=\sum_{n=0}^{\infty}\frac{1}{n!}J_{M1}J_{M2}\cdots J_{Mn}\frac{\delta^n \left\langle \gamma_2,t_2|\gamma_1,t_1\right\rangle}{\delta J_{M1}\delta J_{M2}\cdots \delta J_{Mn}}[J=0]\\
&=\sum_{n=0}^{\infty}\frac{1}{n!}(-i)^nJ_{M1}\cdots J_{Mn}\left\langle \gamma_2,t_2|\mathcal{P}(\lambda^{M1}\cdots \lambda^{Mn})|\gamma_1,t_1\right\rangle[J=0]\\
&=\left\langle \gamma_2,t_2|\mathcal{P}\left\{\mathrm{exp}({-iJ_M\lambda^M}) \right\}|\gamma_1,t_1\right\rangle[J=0].
\end{aligned}
\end{equation}
We acquire the following functional-differential equation by combining Eq. (\ref{B7}) with Eq. (\ref{B8}):
\begin{equation}\label{B9}
\frac{\delta C}{\delta\lambda^M}(J)K_a(\gamma_2,t_2;\gamma_1,t_1)[J]=J_M\left\langle \gamma_2,t_2|\gamma_1,t_1\right\rangle[J],
\end{equation}
whose integration gives rise to the path integral. If we integral over all $\lambda^M$ with boundary conditions $\lambda(\gamma_*;\gamma)\equiv\lambda(\gamma_*;\gamma_1)$ at $t=t_1$ and $\lambda(\gamma_*;\gamma)\equiv\lambda(\gamma_*;\gamma_2)$ at $t=t_2$ to solve Eq. (\ref{B9}), we will find that
\begin{equation}\label{B10}
K_a(\gamma_2,t_2;\gamma_1,t_1)[J]=\int_{\mathcal{M}}\prod_{M}^{\mathrm{dim}(\mathcal{M})}\mathrm{d}\lambda^M\mathrm{e}^{i(C-J_M\lambda^M)},
\end{equation}
where, any vector in $\mathcal{M}$ has ${\mathrm{dim}[\mathcal{M}]}$ components. Changing the integral variables by general rule, namely
\begin{equation}\label{B11}
\begin{aligned}
\prod_{M=1}^{\mathrm{dim}(\mathcal{M})}\mathrm{d}\lambda^M&=\prod_{N=1}^{\mathrm{dim}(\mathcal{M})}\mathrm{d}X^N\left|\mathrm{det}\frac{\delta\lambda^M}{\delta X^N} \right|\\
&=\prod_{M}^{\mathrm{dim}(\mathcal{M})}\mathrm{d}X^M\frac{\sqrt{G_{MN}(\gamma)}}{\sqrt{G_{MN}(\gamma_*)}}|\Delta(\gamma_*;\gamma)|,
\end{aligned}
\end{equation}
where $|\Delta(\gamma_*;\gamma)|$ is defined in Eq. (\ref{3.6}) as the Van Vleck-Morrete determinant. The original Haar measure Eq. (\ref{3.1}) becomes
\begin{equation}\label{B12}
[\mathrm{d}\gamma]\to\frac{N_c|\Delta(\gamma_*;\gamma)|}{\sqrt{G_{MN}(\gamma_*)}}[\mathrm{d}\gamma].
\end{equation}
Assuming that $J=0$ for Eq. (\ref{B10}), we substitute Eq. (\ref{B11}) into Eq. (\ref{3.4}) and ultlize the following property
\begin{equation}\label{B13}
K_a(\gamma_N,t_N;\gamma_0,t_0)=\prod_{i=0}^{N-1}K_a(\gamma_{i+1},t_{i+1};\gamma_{i},t_{i})
\end{equation}
to obtain Eq. (\ref{3.9}) by absorbing all factors into the path integral measure.

\section{Principle of minimal complexity and second law of complexity}\label{F3}
Analogous to the relationship between the least action principle and the maximum entropy suggested by the thermodynamic second law discussed in \cite{86}, we will provide herein remarks on the connection between \textit{the principle of minimal complexity} and \textit{the second law of complexity}. We first need to identify a stationary distribution $P(\gamma)$. According to the principle of Jaynes, the entropy of the auxiliary system (i.e., the Shannon entropy of $\mathcal{A}$) can be maximized under two certain constraints, namely
\begin{equation}\label{FC1}
\begin{aligned}
&\langle C\rangle=\int_{\mathcal{M}} P(\gamma)C(\gamma)D\gamma\quad \mathrm{and}
&\int_{\mathcal{M}} P(\gamma)D\gamma=\mathrm{Constant}
\end{aligned}
\end{equation}
to obtain the optimal distribution, where the first constraint is satisfied by the ergodicity and the second constraint is a normalized condition. The entropy of $\mathcal{A}$ is denoted by symbol $S_a$. Using the Lagrange multiplier method to maximize $S_a$, we obtain
\begin{equation}\label{FC2}
\delta \left[-S_a+\alpha\int_M P(\gamma)D\gamma+\eta\int_{\mathcal{M}} P(\gamma)C(\gamma)D\gamma \right]=0,
\end{equation}
where $\alpha$ and $\eta$ are Lagrangian multipliers\footnote{The stationary distribution requires $\eta$ to be strictly greater than zero and a constant \cite{85}.}. The entropy is given as
\begin{equation}\label{FC3}
S_a\equiv -\int_{\mathcal{M}}P(\gamma)\mathrm{log}P(\gamma)D\gamma,
\end{equation}
where $\eta$ is a positive constant number. The following distribution is stationary, namely Eq. (\ref{3.11}). By substituting the stationary distribution into the Eq. (\ref{FC3}), we find that
\begin{equation}\label{FC4}
S_a\propto \langle C\rangle
\end{equation}
because $\mathcal{N}_a$ and $\eta$ are constants. The average complexity is maximized when system $\mathcal{A}$ reaches its thermal equilibrium (corresponding to the state with the largest $S_a$). This is in line with the description of \textit{the second law of complexity} \cite{8}.

We now focus on \textit{the principle of minimal complexity}. We consider the perturbation of the distribution as $\delta P(C(\gamma))$; however, we obtain the following because $P(C(\gamma))$ is stationary, such that $\delta P(C(\gamma))=0$:
\begin{equation}\label{FC5}
\delta P(C(\gamma))=-\eta P(C(\gamma))\delta C(\gamma)=0.
\end{equation}
$\delta C=0$ gives rise to the Euler-Lagrange equation, that is, Eq. (\ref{2.19}), which meets the requirement of \textit{the principle of minimal complexity}. Fig. 7 illustrates the core idea.

\begin{figure}[htbp]
\centering
\includegraphics[scale=0.5]{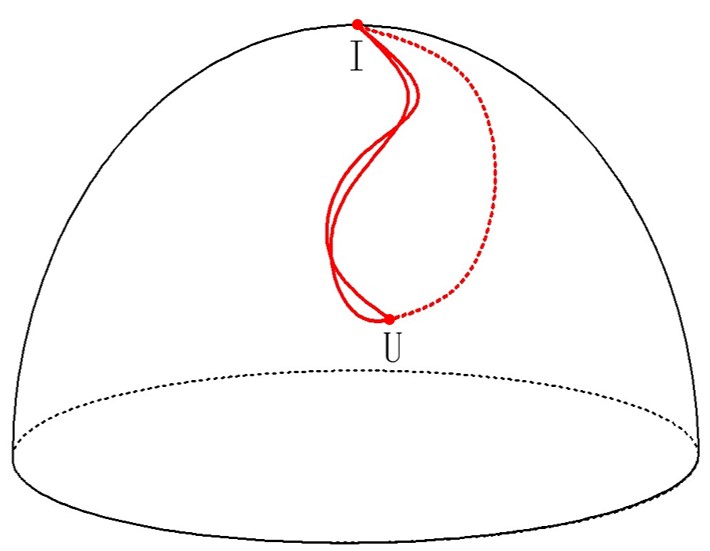} \label{PF1}
\caption{Only the trajectories with minimal complexity (denoted by solid lines) contribute to the path integral, the contribution of the trajectory with large complexity (denoted by red dotted line) is exponentially suppressed by the factor $e^{-\eta C}$ in Eq. (\ref{3.9}).}
\end{figure}
Combining the above discussions, we learned that averaging over the trajectories satisfying \textit{the principle of minimal complexity} directly produces the maximum average complexity based on \textit{the second law of complexity}. One may argue that we can only identify trajectories with extreme values from the first-order variation that $\delta C=0$ but not the minimal values. However, since the probability of each trajectory is weighted by a factor $\mathrm{e}^{-\eta C}$ in Eq. (\ref{4.7}), when a trajectory's complexity is large, its contribution is almost negligible. Then, these trajectories satisfy \textit{the principle of minimal complexity}. In other words, we do not need to test the second-order variation to find the minimal complexity in the average sense.

\section{Glossary and some clarifications}\label{F5}
We provide some supplementary explainations to the text here to help readers clearly understand the content of this paper.

\textit{Notations}: The uppercase letters $Y$, $M$, $N$, and $S$ used in this paper denote the components of vector $\gamma\in\mathcal{M}$. $x^i$ and $x^j$ represent the components of vector $\vec{x}$. Subscript $k$, $i$, and $j$ in Section \ref{C44} and Eq. (\ref{2.22}) depict the sites $k$, $i$, and $j$ affected by the local gate. In Section \ref{C5}, subscript $k$ indicates the $k$ momentum sector.

\textit{Generalized Pauli matrices}: The generalized Pauli matrices throughout this paper have the same meaning as in \cite{8}, specifically footnote 5 of \cite{8}.

\textit{$k$-local Hamiltonian}: The $k$-local Hamiltonian only contains terms acting on $k$ or fewer qubits. For example, Eq. (\ref{5.1}) is a $2$-local Hamiltonian.

\textit{Irreducible Markov chain}: A Markov chain is irreducible if a chain of steps exists between any two states with a positive probability \cite{88}.

\textit{Relationship between complexity and entropy}: This relationship is explored in \cite{8,67}. The content mainly includes the following points:
\begin{itemize}
\item[$\bullet$]
The entropy of the auxiliary system $\mathcal{A}$ is proportional to the ensemble-averaged complexity of quantum system $\mathcal{Q}$.
\end{itemize}
\begin{itemize}
\item[$\bullet$]
The physical law of the complexity operates on a vastly longer time than the entropy, such as the recurrence phenomenon.
\end{itemize}
\begin{itemize}
\item[$\bullet$]
The complexity along the particle trajectory is equal to the entropy of a black hole in Rindler units if the energy of the particle in the system $\mathcal{A}$ is conserved.
\end{itemize}

\textit{Constant $\eta$ and computational free energy $F_a$}: In Section \ref{C321}, we mentioned that $\eta$ can be seen as the inverse temperature $1/T_a$ of system $\mathcal{A}$ or a positive Lagrangian multiplier (Appendix \ref{F3}). By considering $\eta$ as a small inverse temperature, Eq. (\ref{3.20}) represents the thermodynamic free energy of the system $\mathcal{A}$. In \cite{29}, Feynman has discussed the relationship between thermodynamic free energy and the free energy obtained through path integral. To better understand the point, we assume that $\mathcal{M}=\mathrm{SU}(2)$, which corresponds to a nonrelativistic free particle moving in a three-dimensional (3D) configuration space, and choose the standard inner-product metric instead of the complexity metric for simplicity. The partition function\footnote{The general expression of thermodynamic partition function of a three-dimensional free particle reads: $Z^{th}=\left((mk_BT_{\mathrm{th}})/(2\pi\hbar^2)\right)^{\frac{3}{2}}$, where $m$, $k_B$ and $T_{\mathrm{th}}$ represent the mass of the particle, Boltzmann constant and the thermodynamic temperature, respectively. However, for simplicity we set $\hbar=1$ and $k_B=1$ in our work.} is directly written as follows:
\begin{equation}\label{F51}
Z_a^{\mathrm{th}}=\left(\frac{T_a}{2\pi}\right)^{\frac{3}{2}}.
\end{equation}
The metric we chose is the standard inner-product metric that forms an Euclidean space. Hence, we take the Gaussian integral of Eq. (\ref{3.9}), which is the path integral in 3D Euclidean space \cite{58}, to give
\begin{equation}\label{F52}
Z_a=\left(\frac{1}{2\pi\eta T}\right)^{\frac{3}{2}},
\end{equation}
where $T$ denotes the time interval. If we let $\eta=1/T_a$ and set $T=1$, Eq. (\ref{F52})  then coincides with $Z_a^{\mathrm{th}}$. Accordingly, the computational free energy $F_a$ obtained from Eq. (\ref{F52}) is equivalent to the thermodynamic free energy of the system $\mathcal{A}$. Consequently, from Eq. (\ref{4.25}) the computational free energy difference $-\Delta F_a(T)=F_a(0)-F_a(T)$ can be explained as the possible maximal value of the average computational work available for extraction by the system $\mathcal{A}$ in a certain time interval $T=1$. For such a perspective, the temperature of the system $\mathcal{A}$ is considered to be determined in several ways \cite{8}. The simplest choice presented in \cite{8} only depends on the number of qubits and the locality parameter $k$ ($k$-local):
\begin{equation}\label{F53}
T_a\propto\frac{1}{K^{k-1}}.
\end{equation}
Note that $T_a$ is the temperature of system $\mathcal{A}$, but not the temperature of quantum system $\mathcal{Q}$.

\textit{Boltzmann and Gaussian distributions}: Boltzmann distributions with a quadratic Hamiltonian are also Gaussian distributions by considering the velocity as a stochastic variable.

\textit{Quasi-static limit}: In Section \ref{C5}, $T\to \infty$ corresponds to the reversible limit, which is similar to the ``quasi-static'' limit in thermodynamics. Discussing the thermodynamics of the auxiliary system $\mathcal{A}$, this limit can be regarded as the quasi-static limit of system $\mathcal{A}$. Note that we can only talk about such a limit for the classical system $\mathcal{A}$ but not for the quantum system $\mathcal{Q}$.

\end{document}